\pdfoutput=1

\documentclass[12pt,a4paper]{article}

\usepackage{ifthen} 
\newboolean{pdflatex}
\setboolean{pdflatex}{true} 

\newboolean{articletitles}
\setboolean{articletitles}{true} 

\newboolean{uprightparticles}
\setboolean{uprightparticles}{false} 

\newboolean{inbibliography}
\setboolean{inbibliography}{false} 

\usepackage{booktabs}
\usepackage{microtype}
\usepackage{lineno}  
\usepackage{xspace} 
\usepackage{caption} 

\usepackage{graphicx}  
\usepackage{color}
\usepackage{colortbl}

\usepackage{amsmath} 
\usepackage{amssymb}
\usepackage{amsfonts}
\usepackage{upgreek} 

\newcommand*\patchAmsMathEnvironmentForLineno[1]{%
\expandafter\let\csname old#1\expandafter\endcsname\csname #1\endcsname
\expandafter\let\csname oldend#1\expandafter\endcsname\csname
end#1\endcsname
 \renewenvironment{#1}%
   {\linenomath\csname old#1\endcsname}%
   {\csname oldend#1\endcsname\endlinenomath}%
}
\newcommand*\patchBothAmsMathEnvironmentsForLineno[1]{%
  \patchAmsMathEnvironmentForLineno{#1}%
  \patchAmsMathEnvironmentForLineno{#1*}%
}
\AtBeginDocument{%
\patchBothAmsMathEnvironmentsForLineno{equation}%
\patchBothAmsMathEnvironmentsForLineno{align}%
\patchBothAmsMathEnvironmentsForLineno{flalign}%
\patchBothAmsMathEnvironmentsForLineno{alignat}%
\patchBothAmsMathEnvironmentsForLineno{gather}%
\patchBothAmsMathEnvironmentsForLineno{multline}%
\patchBothAmsMathEnvironmentsForLineno{eqnarray}%
}

\usepackage{hyperref}    
\usepackage[all]{hypcap} 

\def\lhcb {\mbox{LHCb}\xspace}

\def\babar  {\mbox{BaBar}\xspace}
\def\belle  {\mbox{Belle}\xspace}

\ifthenelse{\boolean{uprightparticles}}%
{
 
 \def\Pgamma      {\ensuremath{\upgamma}\xspace}

 \def\Ppi         {\ensuremath{\uppi}\xspace}                 
                  
 \def\Prho        {\ensuremath{\uprho}\xspace}

 \def\Ppsi        {\ensuremath{\uppsi}\xspace}

 \def\PDelta      {\ensuremath{\Delta}\xspace}                 
 \def\PXi      {\ensuremath{\Xi}\xspace}                 
 \def\PLambda      {\ensuremath{\Lambda}\xspace}                 
 \def\PSigma      {\ensuremath{\Sigma}\xspace}                 
 \def\POmega      {\ensuremath{\Omega}\xspace}                 
 \def\PUpsilon      {\ensuremath{\Upsilon}\xspace}                 
 
 \def\PB      {\ensuremath{\mathrm{B}}\xspace}                 
                  
 \def\PD      {\ensuremath{\mathrm{D}}\xspace}

 \def\PJ      {\ensuremath{\mathrm{J}}\xspace}                 
 \def\PK      {\ensuremath{\mathrm{K}}\xspace}

 \def\PW      {\ensuremath{\mathrm{W}}\xspace}

 \def\Pb      {\ensuremath{\mathrm{b}}\xspace}                 
 \def\Pc      {\ensuremath{\mathrm{c}}\xspace}

 \def\Pi      {\ensuremath{\mathrm{i}}\xspace}

 \def\Ps      {\ensuremath{\mathrm{s}}\xspace}                 
                  
 \def\Pu      {\ensuremath{\mathrm{u}}\xspace}

}
{
 
 \def\Pgamma      {\ensuremath{\gamma}\xspace}

 \def\Ppi         {\ensuremath{\pi}\xspace}                 
                  
 \def\Prho        {\ensuremath{\rho}\xspace}

 \def\Ppsi        {\ensuremath{\psi}\xspace}                 
                  
 \mathchardef\PDelta="7101
 \mathchardef\PXi="7104
 \mathchardef\PLambda="7103
 \mathchardef\PSigma="7106
 \mathchardef\POmega="710A
 \mathchardef\PUpsilon="7107
                  
 \def\PB      {\ensuremath{B}\xspace}                 
                  
 \def\PD      {\ensuremath{D}\xspace}

 \def\PJ      {\ensuremath{J}\xspace}                 
 \def\PK      {\ensuremath{K}\xspace}

 \def\PW      {\ensuremath{W}\xspace}

 \def\Pb      {\ensuremath{b}\xspace}                 
 \def\Pc      {\ensuremath{c}\xspace}

 \def\Pi      {\ensuremath{i}\xspace}

 \def\Ps      {\ensuremath{s}\xspace}                 
                  
 \def\Pu      {\ensuremath{u}\xspace}

}

\makeatletter
\ifcase \@ptsize \relax
  \newcommand{\miniscule}{\@setfontsize\miniscule{4}{5}}
\or
  \newcommand{\miniscule}{\@setfontsize\miniscule{5}{6}}
\or
  \newcommand{\miniscule}{\@setfontsize\miniscule{5}{6}}
\fi
\makeatother

\DeclareRobustCommand{\optbar}[1]{\shortstack{{\miniscule (\rule[.5ex]{1.25em}{.18mm})}
  \\ [-.7ex] $#1$}}

\def\g      {{\ensuremath{\Pgamma}}\xspace}

\def\W      {{\ensuremath{\PW}}\xspace}

\def\uquark    {{\ensuremath{\Pu}}\xspace}

\def\squark    {{\ensuremath{\Ps}}\xspace}

\def\cquark    {{\ensuremath{\Pc}}\xspace}

\def\bquark    {{\ensuremath{\Pb}}\xspace}

\def\pion   {{\ensuremath{\Ppi}}\xspace}
\def\piz    {{\ensuremath{\pion^0}}\xspace}

\def\pip    {{\ensuremath{\pion^+}}\xspace}
\def\pim    {{\ensuremath{\pion^-}}\xspace}
\def\pipm   {{\ensuremath{\pion^\pm}}\xspace}
\def\pimp   {{\ensuremath{\pion^\mp}}\xspace}

\def\rhomeson {{\ensuremath{\Prho}}\xspace}
\def\rhoz     {{\ensuremath{\rhomeson^0}}\xspace}

\def\kaon    {{\ensuremath{\PK}}\xspace}
  \def\Kbar    {{\kern 0.2em\overline{\kern -0.2em \PK}{}}\xspace}

\def\KorKbar    {\kern 0.18em\optbar{\kern -0.18em K}{}\xspace}

\def\Kp      {{\ensuremath{\kaon^+}}\xspace}
\def\Km      {{\ensuremath{\kaon^-}}\xspace}
\def\Kpm     {{\ensuremath{\kaon^\pm}}\xspace}
\def\Kmp     {{\ensuremath{\kaon^\mp}}\xspace}

\def\Kstarz  {{\ensuremath{\kaon^{*0}}}\xspace}
\def\Kstarzb {{\ensuremath{\Kbar{}^{*0}}}\xspace}
\def\Kstar   {{\ensuremath{\kaon^*}}\xspace}

  \def\Dbar    {{\kern 0.2em\overline{\kern -0.2em \PD}{}}\xspace}
\def\D       {{\ensuremath{\PD}}\xspace}

\def\DorDbar    {\kern 0.18em\optbar{\kern -0.18em D}{}\xspace}
\def\Dz      {{\ensuremath{\D^0}}\xspace}
\def\Dzb     {{\ensuremath{\Dbar{}^0}}\xspace}
\def\Dp      {{\ensuremath{\D^+}}\xspace}

\def\Dstar   {{\ensuremath{\D^*}}\xspace}

\def\Dstarz  {{\ensuremath{\D^{*0}}}\xspace}
\def\Dstarzb {{\ensuremath{\Dbar{}^{*0}}}\xspace}

\def\Dstarpm {{\ensuremath{\D^{*\pm}}}\xspace}

\def\Ds      {{\ensuremath{\D^+_\squark}}\xspace}

\def\B       {{\ensuremath{\PB}}\xspace}
\def\Bbar    {{\ensuremath{\kern 0.18em\overline{\kern -0.18em \PB}{}}}\xspace}
\def\Bb      {{\ensuremath{\Bbar}}\xspace}
\def\BorBbar    {\kern 0.18em\optbar{\kern -0.18em B}{}\xspace}
\def\Bz      {{\ensuremath{\B^0}}\xspace}
\def\Bzb     {{\ensuremath{\Bbar{}^0}}\xspace}
\def\Bu      {{\ensuremath{\B^+}}\xspace}

\def\Bp      {{\ensuremath{\Bu}}\xspace}

\def\Bpm     {{\ensuremath{\B^\pm}}\xspace}

\def\Bd      {{\ensuremath{\B^0}}\xspace}
\def\Bs      {{\ensuremath{\B^0_\squark}}\xspace}
\def\Bsb     {{\ensuremath{\Bbar{}^0_\squark}}\xspace}

\def\jpsi     {{\ensuremath{{\PJ\mskip -3mu/\mskip -2mu\Ppsi\mskip 2mu}}}\xspace}

  \def\Y#1S{\ensuremath{\PUpsilon{(#1S)}}\xspace}

\def\Lz          {{\ensuremath{\PLambda}}\xspace}
\def\Lbar        {{\ensuremath{\kern 0.1em\overline{\kern -0.1em\PLambda}}}\xspace}
\def\LorLbar    {\kern 0.18em\optbar{\kern -0.18em \PLambda}{}\xspace}

\def\Lb      {{\ensuremath{\Lz^0_\bquark}}\xspace}

\def\BF         {{\ensuremath{\cal B}}\xspace}

\def\BR         {\BF}

\def\to                 {\ensuremath{\rightarrow}\xspace}

\def\CP                {{\ensuremath{C\!P}}\xspace}

\def\AT#1     {\ensuremath{A_{\mathrm{T}}^{#1}}\xspace}           

\def\C#1      {\ensuremath{\mathcal{C}_{#1}}\xspace}                       
\def\Cp#1     {\ensuremath{\mathcal{C}_{#1}^{'}}\xspace}                    
\def\Ceff#1   {\ensuremath{\mathcal{C}_{#1}^{\mathrm{(eff)}}}\xspace}        
\def\Cpeff#1  {\ensuremath{\mathcal{C}_{#1}^{'\mathrm{(eff)}}}\xspace}       
\def\Ope#1    {\ensuremath{\mathcal{O}_{#1}}\xspace}                       
\def\Opep#1   {\ensuremath{\mathcal{O}_{#1}^{'}}\xspace}                    

\newcommand{\tev}{\ifthenelse{\boolean{inbibliography}}{\ensuremath{~T\kern -0.05em eV}\xspace}{\ensuremath{\mathrm{\,Te\kern -0.1em V}}}\xspace}
\newcommand{\gev}{\ensuremath{\mathrm{\,Ge\kern -0.1em V}}\xspace}
\newcommand{\mev}{\ensuremath{\mathrm{\,Me\kern -0.1em V}}\xspace}
\newcommand{\kev}{\ensuremath{\mathrm{\,ke\kern -0.1em V}}\xspace}
\newcommand{\ev}{\ensuremath{\mathrm{\,e\kern -0.1em V}}\xspace}
\newcommand{\gevc}{\ensuremath{{\mathrm{\,Ge\kern -0.1em V\!/}c}}\xspace}
\newcommand{\mevc}{\ensuremath{{\mathrm{\,Me\kern -0.1em V\!/}c}}\xspace}
\newcommand{\gevcc}{\ensuremath{{\mathrm{\,Ge\kern -0.1em V\!/}c^2}}\xspace}
\newcommand{\gevgevcccc}{\ensuremath{{\mathrm{\,Ge\kern -0.1em V^2\!/}c^4}}\xspace}
\newcommand{\mevcc}{\ensuremath{{\mathrm{\,Me\kern -0.1em V\!/}c^2}}\xspace}

\def\mum  {\ensuremath{{\,\upmu\rm m}}\xspace}

\def\invfb   {\ensuremath{\mbox{\,fb}^{-1}}\xspace}

\newcommand{\chisq}{\ensuremath{\chi^2}\xspace}

\newcommand{\chisqip}{\ensuremath{\chi^2_{\rm IP}}\xspace}

\def\gsim{{~\raise.15em\hbox{$>$}\kern-.85em
          \lower.35em\hbox{$\sim$}~}\xspace}
\def\lsim{{~\raise.15em\hbox{$<$}\kern-.85em
          \lower.35em\hbox{$\sim$}~}\xspace}

\def\PDF {PDF\xspace}

\def\sPlot{\mbox{\em sPlot}}

\def\ptot       {\mbox{$p$}\xspace}
\def\pt         {\mbox{$p_{\rm T}$}\xspace}

\def\evtgen     {\mbox{\textsc{EvtGen}}\xspace}

\def\geant      {\mbox{\textsc{Geant4}}\xspace}

\def\photos     {\mbox{\textsc{Photos}}\xspace}

\def\pythia     {\mbox{\textsc{Pythia}}\xspace}

\def\tell1  {TELL1\xspace}
\def\ukl1   {UKL1\xspace}

\newcommand{\ie}{\mbox{\itshape i.e.}\xspace}

\usepackage{cite} 
\usepackage{mciteplus}
\usepackage{cleveref}

\crefname{equation}{Eq.}{Eqs.}
\crefrangelabelformat{equation}{(#3#1#4--#5#2#6)}

\usepackage{longtable} 

\begin{document}

\renewcommand{\thefootnote}{\fnsymbol{footnote}}
\setcounter{footnote}{1}


\begin{titlepage}
\pagenumbering{roman}

\vspace*{-1.5cm}
\centerline{\large EUROPEAN ORGANIZATION FOR NUCLEAR RESEARCH (CERN)}
\vspace*{1.5cm}
\hspace*{-0.5cm}
\begin{tabular*}{\linewidth}{lc@{\extracolsep{\fill}}r}
\ifthenelse{\boolean{pdflatex}}
{\vspace*{-2.7cm}\mbox{\!\!\!\includegraphics[width=.14\textwidth]{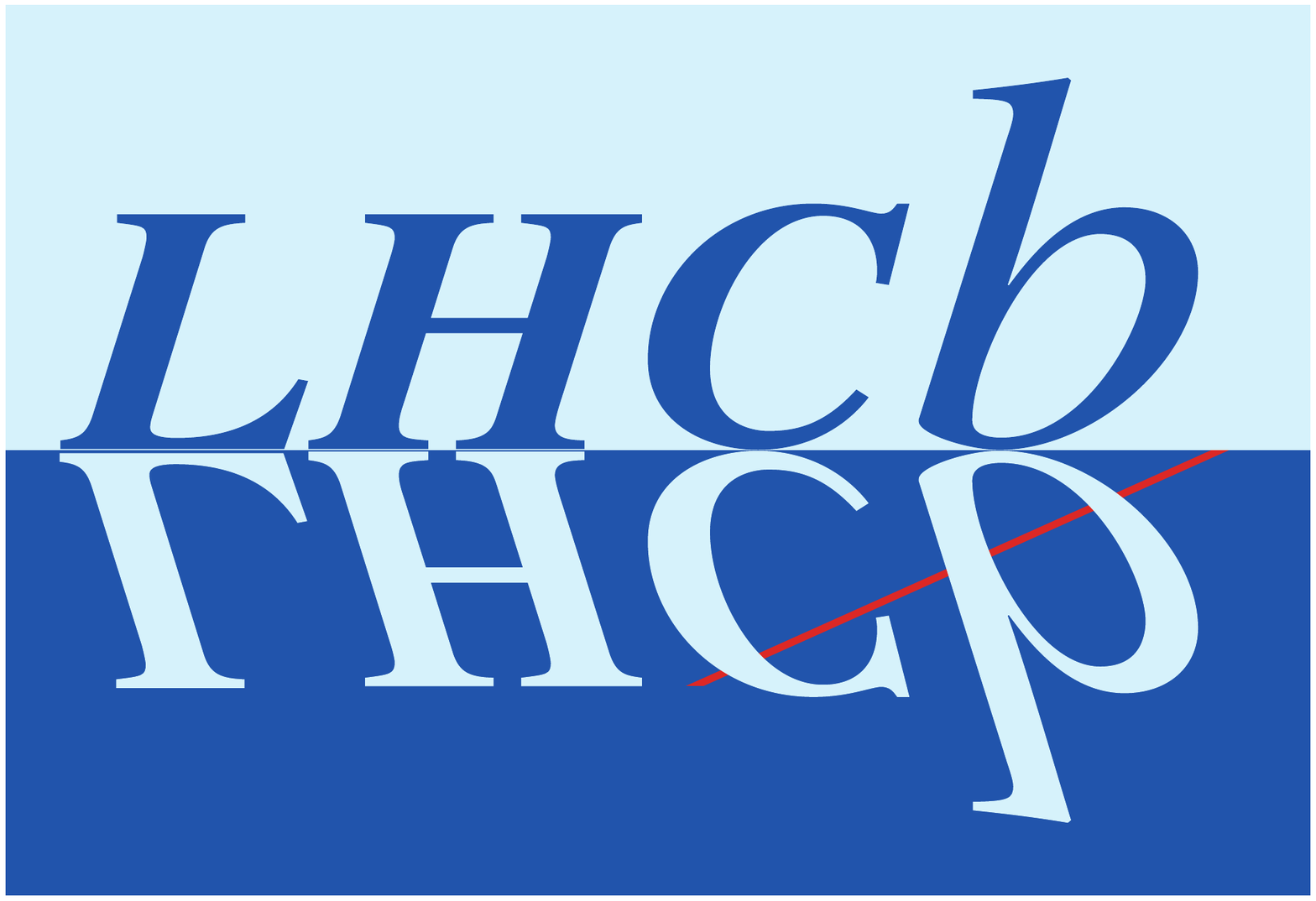}} & &}%
{\vspace*{-1.2cm}\mbox{\!\!\!\includegraphics[width=.12\textwidth]{lhcb-logo.eps}} & &}%
\\
 & & CERN-PH-EP-2014-182 \\  
 & & LHCb-PAPER-2014-028 \\  
 & & 2 December 2014 \\ 
 & & \\
\end{tabular*}

\vspace*{4.0cm}

{\bf\boldmath\huge
\begin{center}
Measurement of \CP violation parameters in $\Bz \to \D \Kstarz$ decays
\end{center}
}

\vspace*{2.0cm}

\begin{center}
The LHCb collaboration\footnote{Authors are listed at the end of this paper.}
\end{center}

\vspace{\fill}

\begin{abstract}
  \noindent
  An analysis of $\Bz\to\D\Kstarz$ decays is presented, where \D represents an admixture of \Dz and \Dzb mesons reconstructed in four 
separate final states: $\Km\pip$, $\pim\Kp$, $\Kp\Km$ and $\pip\pim$. The data sample corresponds
  to 3.0\invfb of proton-proton collision, collected by the \lhcb 
experiment. 
Measurements of several observables are performed, including \CP asymmetries. The most precise determination is presented
of $r_{\B}(\D\Kstarz)$, the magnitude of the ratio of the amplitudes of the decay $\Bz \to \D\Kp\pim$ with a $\bquark\to\uquark$
or a $\bquark\to\cquark$ transition, 
in a $K\pi$ mass region of $\pm50\mevcc$ around the $K^*(892)$ mass and for an absolute value of the cosine of the \Kstarz helicity angle larger than 0.4.
\end{abstract}

\vspace*{2.0cm}

\begin{center}
  Published in Phys.~Rev.~{\bf D90} (2014) 112002 
\end{center}

\vspace{\fill}

{\footnotesize 
\centerline{\copyright~CERN on behalf of the \lhcb collaboration, license \href{http://creativecommons.org/licenses/by/4.0/}{CC-BY-4.0}.}}
\vspace*{2mm}

\end{titlepage}


\newpage
\setcounter{page}{2}
\mbox{~}

\cleardoublepage


\renewcommand{\thefootnote}{\arabic{footnote}}
\setcounter{footnote}{0}



\pagestyle{plain} 
\setcounter{page}{1}
\pagenumbering{arabic}


%

\section{Introduction}
\label{Introduction}

\begin{figure}[!b]
\centering
\includegraphics[width=0.9\textwidth]{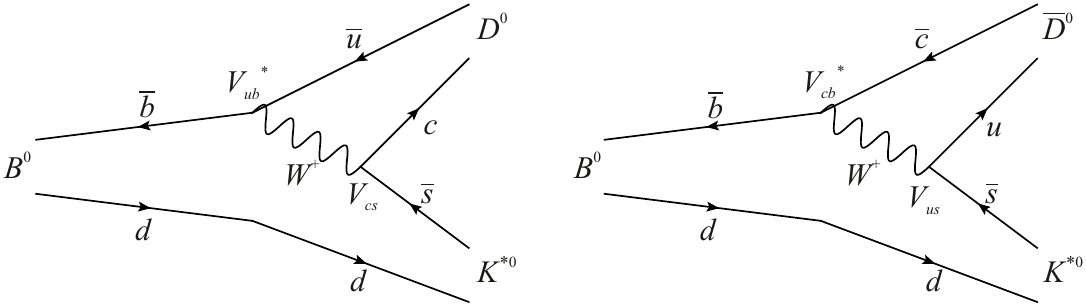}
%
%
%

%
\caption{\small Feynman diagrams of {\it(left)} $\Bz \to \Dz\Kstarz$ and {\it (right)} $\Bz \to \Dzb\Kstarz$.}
\label{fig:feynman_diag}
\end{figure}

Direct \CP violation can arise in $\Bz\to \D \Kstarz$ 
decays from the interference between the two colour-suppressed
$\bquark\to \uquark$  and $\bquark\to \cquark$ transitions shown in the Feynman diagrams of Fig.~\ref{fig:feynman_diag}, when
the \Dz and \Dzb mesons decay to a common final state. Here and in the following, \D represents a neutral meson that is an admixture 
of \Dz and \Dzb mesons and \Kstarz represents the $K^*(892)^0$ meson. Inclusion of 
charge conjugate processes is implied unless specified otherwise.

The amount of \CP violation is related to the value of the weak phase 
\begin{equation}
\gamma \equiv \arg\left(- \frac{V_{ud}V_{ub}^*}{V_{cd}V_{cb}^*}\right),
\end{equation} 
the least-well determined angle of the unitarity triangle, where $V_{ij}$ are elements of the Cabibbo-Kobayashi-Maskawa 
(CKM) matrix~\cite{Cabibbo:1963yz,*Kobayashi:1973fv}. The current experimental measurements are 
$\gamma=\left(72.0_{-15.6}^{+14.7}\right)^\circ$ by the LHCb collaboration~\cite{LHCb-PAPER-2013-020}, 
$\gamma=\left(69_{-16}^{+17}\right)^\circ$ by the \babar~\cite{Lees:2013nha}
collaboration and $\gamma=\left(68_{-14}^{+15}\right)^\circ$ by the \belle collaboration~\cite{Trabelsi:2013uj}.
This angle can be measured with extremely small theoretical uncertainties~\cite{Brod:2013sga}, using decay modes proceeding 
through amplitudes involving only the exchange of a \W boson. 
Such methods to determine $\gamma$ from hadronic \B-decay rates were originally proposed in Refs.~\cite{bib:GW,*bib:GL,*Atwood:1996ci,Atwood:2000ck} 
for $\B\to \D\kaon$ decays and can be applied to the $\Bz\to \D\Kstarz$ decay~\cite{Dunietz:1991yd}. 
In this decay, the charge of the kaon from the $\Kstarz\to\Kp\pim$ decay 
unambiguously identifies the flavour of the decaying \B meson. Hence, no flavour tagging is needed.

The use of these specific neutral \B meson decays is interesting since the interfering amplitudes are of comparable size, as opposed to the 
charged $\Bp\to \D\Kp$ decay that involves both colour-suppressed and colour-allowed amplitudes; hence the system 
could exhibit larger \CP-violating effects.
Contributions from \Bz decays to the  $\D\Kp\pim$ final state through non-$\Kstarz$ intermediate
resonances can pollute the $\D\Kstarz$ reconstructed signal candidates because of the large natural width of the $\Kstarz$. 
They are treated following Ref.~\cite{Gronau:2002mu}, with the use of a coherence factor, $\kappa$, in addition to the
hadronic parameters $r_B$ and $\delta_B$, defined as

\begin{align}
\kappa & \equiv \left | \frac{\int \left |A_{cb}(p) A_{ub}(p)\right | e^{i\delta(p)}{\rm d}p}{\sqrt{\int \left | A_{ub}(p) \right |^2{\rm d}p \int \left | A_{cb}(p) \right |^2{\rm d}p}}\right |, \label{kappa_1}\\
\delta_B  & \equiv \arg \left( \frac{\int \left | A_{cb}(p) A_{ub}(p) \right | e^{i\delta(p)}{\rm d}p}{\sqrt{\int \left | A_{ub}(p) \right |^2 {\rm d}p \int \left | A_{cb}(p) \right|^2{\rm d}p}} \right ), \\
r_B & \equiv \sqrt{\frac{\int \left | A_{ub}(p) \right |^2 {\rm d}p}{\int \left | A_{cb}(p) \right |^2 {\rm d}p}}, \label{kappa_3}  
\end{align}
where $A_{ub}(p)$ and
$A_{cb}(p)$ are the amplitudes of the $\bquark\to\uquark$ and $\bquark\to\cquark$ transitions, 
respectively, to the $\Bz\to D\Kp\pim$ decays, $\delta(p)$ is the strong-phase difference between the two amplitudes and $p$ is a point in the 
three-body phase space of the \Bz meson.
The integrals are defined over the phase space considered here, namely in a $\Kp\pim$ mass range of $\pm50\mevcc$ around the
nominal $\Kstarz$ mass~\cite{PDG2012} and for an absolute value of the cosine of the helicity angle $\theta^*$ greater than 0.4,
where $\theta^*$ is defined as the angle between the $K$ momentum and the opposite of the \B momentum in the \Kstarz rest frame.
The formalism of~(\ref{kappa_1})-(\ref{kappa_3}) applies to the generic three-body 
decay $\Bz\to D \Kp \pim$ with any number of intermediate 
resonances included. The integration range is restricted here to the \Kstarz resonance in order to obtain a large value of the
coherence factor.

This paper presents two measurements of the ratio, ${\cal R}_{\CP+}$,  of flavour-averaged partial widths 
of the $\mbox{\Bz\to\D\Kstarz}$ decay with the \D decaying to a \CP-even eigenstate, 
\begin{equation}
\label{eq:R_CPplus}
{\cal R}_{\CP+}  \equiv  2\times\frac{\Gamma(\Bzb \to \D_{\CP+} \Kstarzb) + \Gamma(\Bz \to \D_{\CP+} \Kstarz)}{\Gamma(\Bzb \to \Dz \Kstarzb) + \Gamma(\Bz \to \Dzb \Kstarz)}.
\end{equation}
The relation above is approximated using specific final states of the \D meson as \mbox{${\cal R}_{\CP+} \approx {\cal R}_d^{hh}$}, neglecting corrections from doubly 
Cabibbo-suppressed $\Dz\to\Kp\pim$ decays, with
\begin{equation}
\label{eq:R_d_hh}
{\cal R}_d^{hh} \equiv \frac{\Gamma(\Bzb \to \D(h^+h^-) \Kstarzb) + \Gamma(\Bz \to \D(h^+h^-) \Kstarz)}{\Gamma(\Bzb \to \D(\Km\pip) \Kstarzb) + \Gamma(\Bz \to \D(\Kp\pim) \Kstarz)}
\times \frac{\BF(\Dz\to\Km\pip)}{\BF(\Dz\to h^+h^-)} ,
\end{equation}
where $h$ represents either a $\pi$ or a $K$ meson. This quantity is related to the $\gamma$ angle and the hadronic parameters by~\cite{delAmoSanchez:2010ji}
\begin{equation}
\label{eq:R_d_hh_1}
{\cal R}_d^{hh} = \frac{1+r_B^2+2r_B\kappa\cos\delta_B\cos\gamma}{1+r_B^2r_D^2+2r_Br_D\kappa\cos(\delta_B-\delta_D)\cos\gamma},
\end{equation}
where $r_D$ and $\delta_D$ are the magnitude of the ratio and the phase difference, respectively, between the amplitudes of the $\Dz\to\Kp\pim$ and
$\Dz\to\Km\pip$ decays. Charm mixing and \CP violation in the decays of \D mesons have an effect on the determination of
$\gamma$~\cite{Atwood:2000ck,Rama:2013voa} but are neglected here because of the large expected value of $r_B$.


Measurements of the \Bzb-\Bz partial decay-rate asymmetry, ${\cal A}_d^{hh}$, using $\D\to h^+h^-$ final states are also presented,
\begin{equation}
\label{eq:A_d_hh}
{\cal A}_d^{hh} \equiv \frac{\Gamma(\Bzb \to \D(h^+h^-) \Kstarzb) - \Gamma(\Bz \to \D(h^+h^-) \Kstarz)}{\Gamma(\Bzb \to D(h^+h^-) \Kstarzb) + \Gamma(\Bz \to \D(h^+h^-) \Kstarz)}
= \frac{2r_B\kappa\sin\delta_B\sin\gamma}{1+r_B^2+2r_B\kappa\cos\delta_B\cos\gamma}.
\end{equation}

The \Bzb-\Bz asymmetry, ${\cal A}_d^{K\pi}$, obtained from the Cabibbo-favoured decay $\Bz\to\D\Kstarz$ with $\mbox{\D\to\Kp\pim}$, where the two kaons from the \D 
and the \Kstarz decay have the same sign, is
\begin{multline}
\label{eq:A_d_KPi}
{\cal A}_d^{K\pi} \equiv \frac{\Gamma(\Bzb \to \D(\Km\pip) \Kstarzb) - \Gamma(\Bz \to \D(\Kp\pim) \Kstarz)}{\Gamma(\Bzb \to \D(\Km\pip) \Kstarzb) + \Gamma(\Bz \to \D(\Kp\pim) \Kstarz)} \\ =\frac{2r_Br_D\kappa\sin(\delta_B-\delta_D)\sin\gamma}{1+r_B^2r_D^2+2r_Br_D\kappa\cos(\delta_B-\delta_D)\cos\gamma}.
\end{multline}

The Cabibbo-suppressed decay $\Bz\to\D\Kstarz$ with $\D\to\pip\Km$, where the two kaons have opposite charge, is studied for the first time by \lhcb. 
The ratios of suppressed $\Bz\to\D(\pip\Km)\Kstarz$ to favoured  $\Bz\to\D(\Kp\pim)\Kstarz$ 
partial widths are measured 
separately for \Bz and \Bzb, and defined as ${\cal R}_d^+$ and ${\cal R}_d^-$, respectively, 
\begin{align}
\label{eq:R_d_plusminus}
{\cal{R}}^{+}_d \equiv & \frac{\Gamma(\Bz\to \D(\pip\Km)\Kstarz)} {\Gamma(\Bz\to \D(\Kp\pim)\Kstarz)} = \frac{r_B^2+r_D^2+2r_Br_D\kappa\cos(\delta_B+\delta_D+\gamma)}{1+r_B^2r_D^2+2r_Br_D\kappa\cos(\delta_B-\delta_D+\gamma)}, \\
{\cal{R}}^{-}_d \equiv & \frac{\Gamma(\Bzb\to \D(\pim\Kp)\Kstarzb)} {\Gamma(\Bzb\to \D(\Km\pip)\Kstarzb)} = \frac{r_B^2+r_D^2+2r_Br_D\kappa\cos(\delta_B+\delta_D-\gamma)}{1+r_B^2r_D^2+2r_Br_D\kappa\cos(\delta_B-\delta_D-\gamma)}.
\label{eq:R_d_plusminus_1}
\end{align}

In $pp$ collisions, \Bs mesons are produced and can decay to the same final state, 
$\Bs\to \D \Kstarzb$~\cite{LHCb-PAPER-2011-008}.
Similar asymmetry observables to those defined above for \Bz mesons are measured with \Bs mesons.
These are the \Bsb-\Bs asymmetry, ${\cal A}_s^{hh}$, obtained from the $\Kp\Km$ and $\pip\pim$  final states of the \D meson, 
\begin{equation}
\label{eq:A_s_hh}
{\cal A}^{hh}_s \equiv \frac{\Gamma(\Bsb \to \D(h^+h^-) \Kstarz) - \Gamma(\Bs \to \D(h^+h^-) \Kstarzb)}{\Gamma(\Bsb \to \D(h^+h^-) \Kstarz) + \Gamma(\Bs \to \D(h^+h^-) \Kstarzb)},
\end{equation}
and the asymmetry, ${\cal A}_s^{\pi K}$, from the Cabibbo-favoured decay $\Bs\to\D(\pim\Kp)\Kstarzb$, where the two kaons have opposite charge, 
\begin{equation}
\label{eq:A_s_piK}
{\cal A}^{\pi K}_s \equiv \frac{\Gamma(\Bsb \to \D(\pip\Km) \Kstarz) - \Gamma(\Bs \to \D(\pim\Kp) \Kstarzb)}{\Gamma(\Bsb \to \D(\pip\Km) \Kstarz) + \Gamma(\Bs \to \D(\pim\Kp) \Kstarzb)}.
\end{equation}

The $\Bs\to\D(\Km\pip)\Kstarzb$ decay, where the two kaons have the same charge, is 
highly suppressed and therefore unobserved with the current data sample. Finally, the ratios of the flavour-averaged partial widths 
of the \Bz and $\Bs$ decays, when the \D meson is reconstructed as $\D\to h^+h^-$, ${\cal{R}}^{hh}_{ds}$, are also considered,
\begin{equation}
\label{eq:R_ds_hh}
{\cal{R}}^{hh}_{ds} \equiv \frac{\Gamma(\Bzb\to D(h^+h^-)\Kstarzb) + \Gamma(\Bz\to D(h^+h^-)\Kstarz)} {\Gamma(\Bsb\to D(h^+h^-)\Kstarz) + \Gamma(\Bs\to D(h^+h^-)\Kstarzb)}.
\end{equation} 

The observables related to \Bs decays could in principle also be used to determine the value of $\gamma$. However, the observables pertaining to \Bz mesons are far more 
sensitive, 
owing to the fact that the ratio of interfering amplitudes is closer to unity. Those related to \Bs mesons are measured and reported in this paper but are not yet precise enough 
to provide any constraint on $\gamma$.

\section{The \lhcb detector, data set and event selection}\label{detector}
\label{sec:selection}

The study reported here is based on a data sample of $pp$ collisions 
obtained from 3.0\invfb of integrated luminosity with the LHCb 
detector~\cite{Alves:2008zz}. The centre-of-mass energy was $7\,{\rm TeV}$ during the year 2011, when approximately $1/3$ of the 
data were collected, and $8\,{\rm TeV}$ during the year 2012.

The \lhcb detector~\cite{Alves:2008zz} is a single-arm forward
spectrometer covering the \mbox{pseudorapidity} range $2<\eta <5$,
designed for the study of particles containing \bquark or \cquark
quarks. The detector includes a high-precision tracking system
consisting of a silicon-strip vertex detector surrounding the $pp$
interaction region~\cite{LHCb-DP-2014-001}, a large-area silicon-strip detector located
upstream of a dipole magnet with a bending power of about
$4{\rm\,Tm}$, and three stations of silicon-strip detectors and straw
drift tubes placed downstream of the magnet.
The tracking system provides a measurement of momentum, \ptot,  with
a relative uncertainty that varies from 0.4\% at low momentum to 0.6\% at 100\gevc.
The minimum distance of a track to a primary vertex, the impact parameter, is measured with a resolution of $(15+29/\pt)\mum$,
where \pt is the component of \ptot transverse to the beam, in \gevc.
Different types of charged hadrons are distinguished using information
from two ring-imaging Cherenkov detectors~\cite{LHCb-DP-2012-003}. Photon, electron and
hadron candidates are identified by a calorimeter system consisting of
scintillating-pad and preshower detectors, an electromagnetic
calorimeter and a hadronic calorimeter. Muons are identified by a
system composed of alternating layers of iron and multiwire
proportional chambers.
The trigger~\cite{LHCb-DP-2012-004} consists of a
hardware stage, based on information from the calorimeter and muon
systems, followed by a software stage, which applies a full event
reconstruction.

The analysis uses events triggered at the hardware level either when one of the charged tracks of the signal
decay gives a large enough energy deposit in the calorimeter system (hadron trigger), or when one of the particles
in the event, not reconstructed as forming the signal candidate, fulfills any trigger requirement (\ie\ mainly events triggered
by one high \pt muon, hadron, photon or electron coming from the decay of the other \B meson in the event). 
The software trigger requires a two-, three- or four-track
  secondary vertex with a large sum of the \pt of
  the charged particles and a significant displacement from the primary $pp$
  interaction vertices~(PVs). At least one charged particle should have $\pt >
  1.7\gevc$ and \chisqip with respect to any
  PV greater than 16, where \chisqip is defined as the
  difference in \chisq of a given PV reconstructed with and
  without the considered particle.
 A multivariate algorithm~\cite{BBDT} is used for
  the identification of secondary vertices consistent with the decay
  of a \bquark hadron.

Approximately 1 million simulated events are used to describe the signal shapes and to compute the efficiencies
when data-driven methods are not available. In the simulation, $pp$ collisions are generated using
\pythia~\cite{Sjostrand:2006za,Sjostrand:2007gs} with a specific \lhcb
configuration~\cite{LHCb-PROC-2010-056}.  Decays of hadronic particles
are described by \evtgen~\cite{Lange:2001uf}, in which final state
radiation is generated using \photos~\cite{Golonka:2005pn}. The
interaction of the generated particles with the detector and its
response are implemented using the \geant
toolkit~\cite{Allison:2006ve, *Agostinelli:2002hh} as described in
Ref.~\cite{LHCb-PROC-2011-006}.

Candidate $\Bz\to\D\Kstarz$ decays are reconstructed in events fulfilling these trigger conditions combining $D$  mesons
reconstructed in the $\Kpm\pimp$, $\Kp\Km$ and $\pip\pim$ decays and $\Kstarz$ mesons reconstructed in the $\Kp\pim$ 
final state.
The invariant masses of the \D and \Kstarz mesons are required to be within 20\mevcc and 50\mevcc of their known 
masses~\cite{PDG2012}, respectively.  The $B$ candidate momentum is refit constraining the mass of the $D$ meson to its known value. 
It is required that $\left|\cos\theta^*\right|>0.4$.


A boosted decision tree~(BDT)~\cite{Breiman} is used with the algorithm described in Ref.~\cite{AdaBoost} to separate signal from 
combinatorial background. Separate BDTs are optimised for $\Kpm\pimp$, $\Kp\Km$ and $\pip\pim$ final states of the \D meson. 
In all cases the samples used to train the 
BDT are fully simulated events for the signal and candidates from the upper sideband of the $B$ mass distribution in data for the background. 
This upper sideband is defined 
as events with a $\D\Kstarz$ invariant mass between 5.8\gevcc and 7\gevcc, lying outside the region used for the fit described in Sect.~\ref{sec:fit}. 
The variables used by the BDT to differentiate signal and background are: the \pt of each particle in the final state;
the fit quality of the \D and \Bz vertices; the \Kstarz, \D and \Bz \chisqip; the angle between the \Bz momentum and the vector from the PV
to the \Bz decay vertex; the significance of the displacement of the four final-state tracks from the PV. 

Thresholds on the BDT classifier are optimised with respect to the signal significance of the \Bz decay modes for the three final states
$\Bz\to\D(\pip\Km)\Kstarz$, $\Bz\to\D(\Kp\Km)\Kstarz$ and $\Bz\to\D(\pip\pim)\Kstarz$, where the significance is defined as 
$S/\sqrt{S+B}$ with $S$ and $B$ the expected number of signal and background candidates. The efficiencies of the selection based on the
BDT output classifier are equal to 69\%, 71\% and 75\% for the $\D\to \Kpm\pimp$, $\D\to \Kp\Km$ and $\D\to\pip\pim$ 
decay channels, respectively.

To improve the purity of the data sample, further selection requirements are made in addition to the BDT. Particle identification (PID) criteria are applied 
and only well identified pions and kaons are retained. The kaon identification efficiency of the PID criteria is equal to $87\%$ with a pion
misidentification rate of 5\%.
Possible contamination from $\Lb \to \Dzb p h^-$ decays is reduced by keeping only kaon candidates incompatible with being a proton.

A potentially significant background is due to events where the $K$ from $\D\to \Kpm \pimp$ decays is misidentified as a $\pi$ and
the $\pi$ is simultaneously mis-identified as a $K$. This causes cross-feed from the favoured $\Bz\to\D(\Kp\pim)\Kstarz$ decay into the suppressed 
$\Bz\to\D(\pip\Km)\Kstarz$ decay. A veto is applied on the \D invariant mass computed with a pion mass assignment for the kaon and a kaon mass assignment 
for the pion. Only candidates for which this invariant mass differs by more than 7\mevcc from the known \Dz mass~\cite{PDG2012} are kept, reducing
this background to a negligible level while keeping $97\%$ of the signal candidates.

Another potential background is due to charmless decays $B^0 \to h^{\pm} h'^{\mp} \Kp \pim$, where $h'$ is also $\pi$ or $K$. It is
removed by requiring the \D flight 
distance with respect to the \B vertex to exceed three times its uncertainty.
Specific peaking backgrounds from $B_{(s)}^0 \to D_{(s)}^\mp h^\pm$ decays are eliminated by
applying a veto on candidates for which the invariant mass of three of the four charged mesons is compatible
within $\pm 15\mevcc$ of the known \Dp or \Ds masses. 
 
After all selections are applied, $0.9\%$ of the events contain more than one signal candidate. Only the candidate with the largest
\B flight distance with respect to the PV, divided by its uncertainty, is retained. In case 
several PVs are reconstructed, the PV with respect to which the \B candidate has the smallest displacement is used.

\begin{figure}[!t]
\centering
\includegraphics[width=0.5\textwidth]{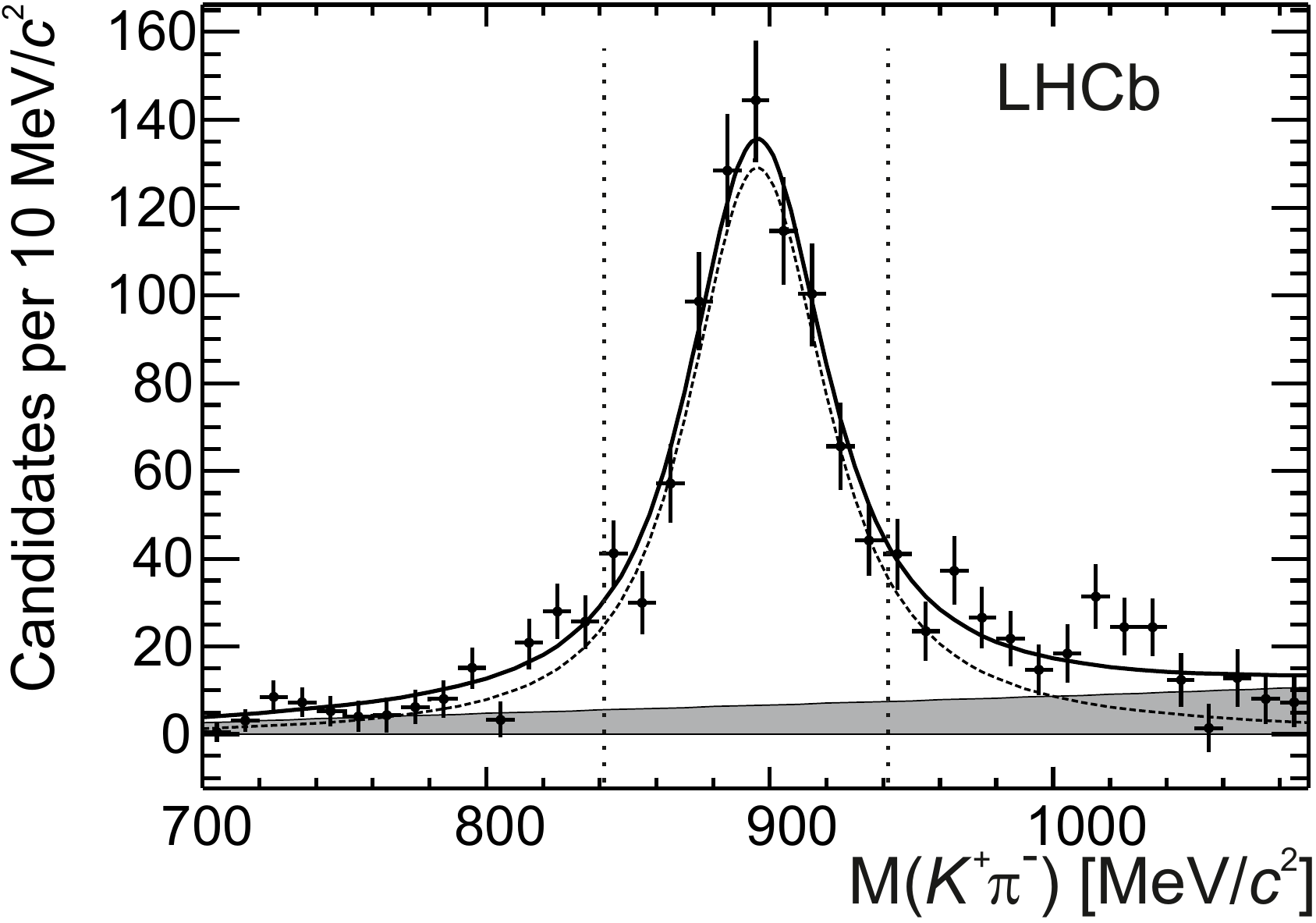}
\caption{\small Background-subtracted $\Kstarz\to\Kp\pim$ invariant mass for $\Bz\to \D(\Kp\pim)\Kstarz$ signal candidates. The data (points) and the fit described in the text (solid line) are shown. The dashed line represents the $\Kstarz$ signal and the filled area the non-\Kstarz contribution to the $\Bz\to\D\Kstarz$ signal. The vertical dotted lines indicate the invariant mass region used in the 
analysis.}
\label{fig:plot_kstar}
\end{figure}

Figure~\ref{fig:plot_kstar} shows the background-subtracted $\Kp\pim$ invariant mass of the \Kstarz candidates used to reconstruct $\Bz\to \D(\Kp\pim)\Kstarz$
decays, obtained with the \sPlot\ technique~\cite{Pivk:2004ty}.
All selections described above have been applied except the requirement on the \Kstarz candidate mass. This distribution is fitted with a relativistic
Breit-Wigner function to describe the \Kstarz signal and a first-order polynomial for the non-\Kstarz contribution. From the fit result, it is estimated
that $(8.4\pm3.4)\%$ of the signal \Bz candidates are formed with a $\Kp\pim$ pair that does not originate from a \Kstarz decay, in the $\Kp\pim$ mass region considered
for the analysis.

\section{Invariant mass fit}
\label{sec:fit}

The numbers of reconstructed signal \Bz and \Bs candidates are determined from an unbinned maximum-likelihood fit to the $D\Kstarz$ 
invariant mass distributions.  
Candidates are split into eight categories, which are fitted simultaneously:
$\D(\Kp\pim) \Kstarz$,
$\D(\Km\pip) \Kstarzb$,
$\D(\pip\Km)\Kstarz$,
$\D(\pim\Kp)\Kstarzb$,
$\D(\Kp\Km) \Kstarz$, 
$\D(\Kp\Km) \Kstarzb$,
$\D(\pip\pim) \Kstarz$ and
$\D(\pip\pim) \Kstarzb$ candidates.
The mass distribution of each category is fitted with a sum of probability density functions 
(PDFs) modelling the various contributing components:

\begin{enumerate}

\item The \Bz and \Bs signals are both described by a sum of two Gaussian functions with a common mean;

\item The combinatorial background is described by an exponential function;

\item The cross-feed from $\Bz\to \D \rho^0$ decays, where one \pion from the $\rho^0\to\pip\pim$ decay
is misidentified as a \kaon, is described by a non-parametric \PDF~\cite{Cranmer:2000du} determined from simulation;

\item The partially reconstructed $\Bz\to \Dstar\Kstarz$ and $\Bsb \to \Dstar \Kstarz$ 
decays, where \Dstar stands for \Dstarz or \Dstarzb with the \piz or $\gamma$ from the $\Dstarz\to\Dz\piz$ or $\Dstarz\to\Dz\gamma$ decay not reconstructed, 
are each modelled by non-parametric PDFs determined from simulation.

\end{enumerate}

A separate fit to $\Bz\to \D(\Kp\pim)\rho^0$ candidates in the same data sample is performed, reconstructing $\rho^0$ in the $\pip\pim$
final state within a $\pm 50\mevcc$ mass range around the known \rhoz mass.
The observed number of $\Bz\to\D(\Kp\pim)\rho^0$ 
candidates is used, along with the efficiency to reconstruct $\Bz\to \D(\Kp\pim)\rho^0$ candidates as $\Bz\to \D(\Kp\pim)\Kstarz$ from simulation, 
to constrain the number of 
cross-feed events in the $\D(\Kp\pim) \Kstarz$ category. 
The numbers of cross-feed candidates in the other categories are derived from the $\D(\Kp\pim) \Kstarz$ category 
using the relative \D branching fractions from Ref.~\cite{PDG2012} 
and selection efficiencies from simulation. 
As a negligible \CP asymmetry is expected for the $\Bz\to \D(\Kp\pim)\rho^0$ background, 
the numbers of cross-feed events in the $\D\Kstarzb$ categories are constrained to be identical 
to those of the corresponding $\D\Kstarz$ categories. 

The partially reconstructed background accumulates at masses lower than the known \Bz mass. 
Its shape depends on the unknown fraction of longitudinal polarisation in the $\Bz\to \Dstar \Kstarz$ 
and $\Bsb\to\Dstar\Kstarz$ decays, \ie the probability that the $\Dstar$ in these decays is produced 
with helicity equal to 0. 
In order to model the $\Bsb\to \Dstar\Kstarz$ contribution, 
a \PDF is built from a linear combination of two non-parametric functions corresponding to the three orthogonal 
helicity amplitudes. Two of the orthogonal helicity amplitudes result in the same distribution in invariant mass because 
of parity conservation in the $\Dstarz\to\Dz\g$ decay, hence simplifying the model.
Each function, modelled from simulated events, corresponds to the weighted sum of the $\Dstarz\to \Dz\gamma$ and $\Dstarz\to \Dz\piz$ contributions 
for a defined helicity eigenstate, where the weights take into account the relative \Dstarz decay branching fractions from Ref.~\cite{PDG2012} and 
the corresponding efficiencies from simulation. The $\Bd\to \Dstar\Kstarz$ background is modelled in a similar way, shifting the shape obtained for 
the $\Bsb\to\Dstar\Kstarz$ decay by the 
known difference between the \Bd and \Bs masses~\cite{PDG2012}.
The coefficients of the two functions in the linear combinations are different for the $\Bz\to\Dstar\Kstarz$ and $\Bsb\to\Dstar\Kstarz$ decays but are 
common to the 8 categories and are free parameters in the fit.

The yields of the \Bs and \Bsb partially reconstructed backgrounds 
in the $\D(\Kp\pim)\Kstarz$ categories are fixed to zero since the $\Bsb\to \Dstar\Kstarz$ decay modes have 
negligible total branching fractions when the kaons from the
\D and \Kstarz have the same charge sign. The yields of the $\Bs\to\Dstar\Kstarzb$ and $\Bsb\to\Dstar\Kstarz$ backgrounds in 
the $\D(\pip\Km)\Kstarz$ categories are constrained to be the same because \CP violation is expected to be negligible 
for this background. Additional constraints on the 
normalisations of the $\Bsb\to\Dstar\Kstarz$ backgrounds in the $\D(\Kp\Km)\Kstarz$ and $\D(\pip\pim)\Kstarz$ categories, relative to the $\D(\pip\Km)\Kstarz$ 
categories, are imposed using the relevant \D decay branching fractions from Ref.~\cite{PDG2012} and selection efficiencies obtained from simulation.

There are 35 free parameters in the fit: the \Bz peak position; the core Gaussian resolution for 
the \Bz and the \Bs signal shapes; the slope of the combinatorial background, which is different for each \D meson final state 
(one parameter for $\D\to \Kpm\pimp$, one for 
$\D\to \Kp\Km$ and one for $\D\to\pip\pim$); the fractions of longitudinal polarisation in the $\Bz \to \Dstar\Kstarz$ and $\Bsb \to \Dstar\Kstarz$ 
backgrounds and the yields for each fit 
component within each category. \CP violation in $\Bz\to\Dstar\Kstarz$ decays is allowed by floating the yields of this background in the $\D\Kstarz$ and 
$\D\Kstarzb$ categories separately. The difference between the central value of the \Bs and \Bz mass is fixed to its known value from Ref.~\cite{PDG2012}
and the ratio between the signal Gaussian resolutions is fixed from the simulation.

\begin{figure}[!b]
\centering
\includegraphics[width=0.49\textwidth]{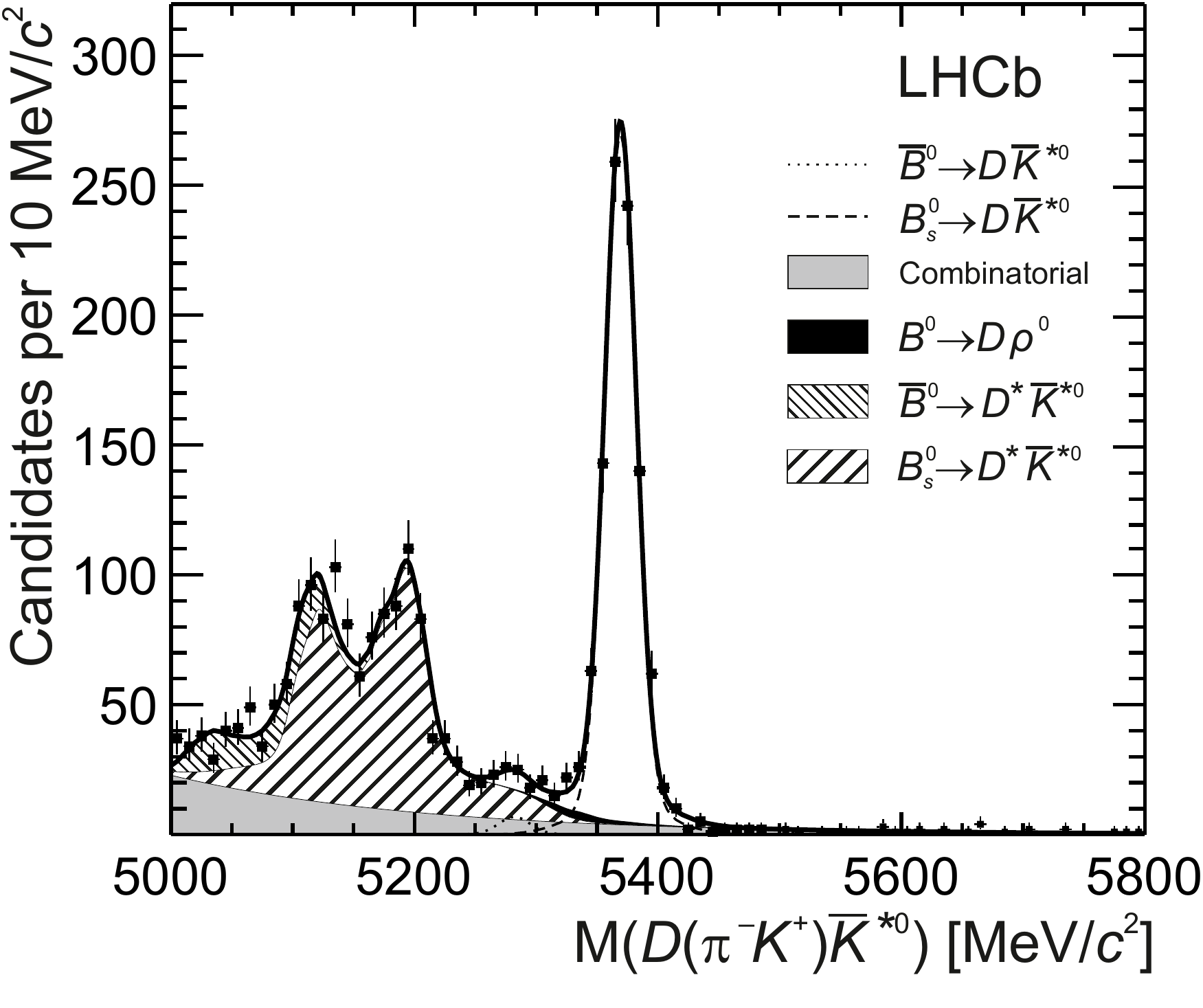}
\includegraphics[width=0.49\textwidth]{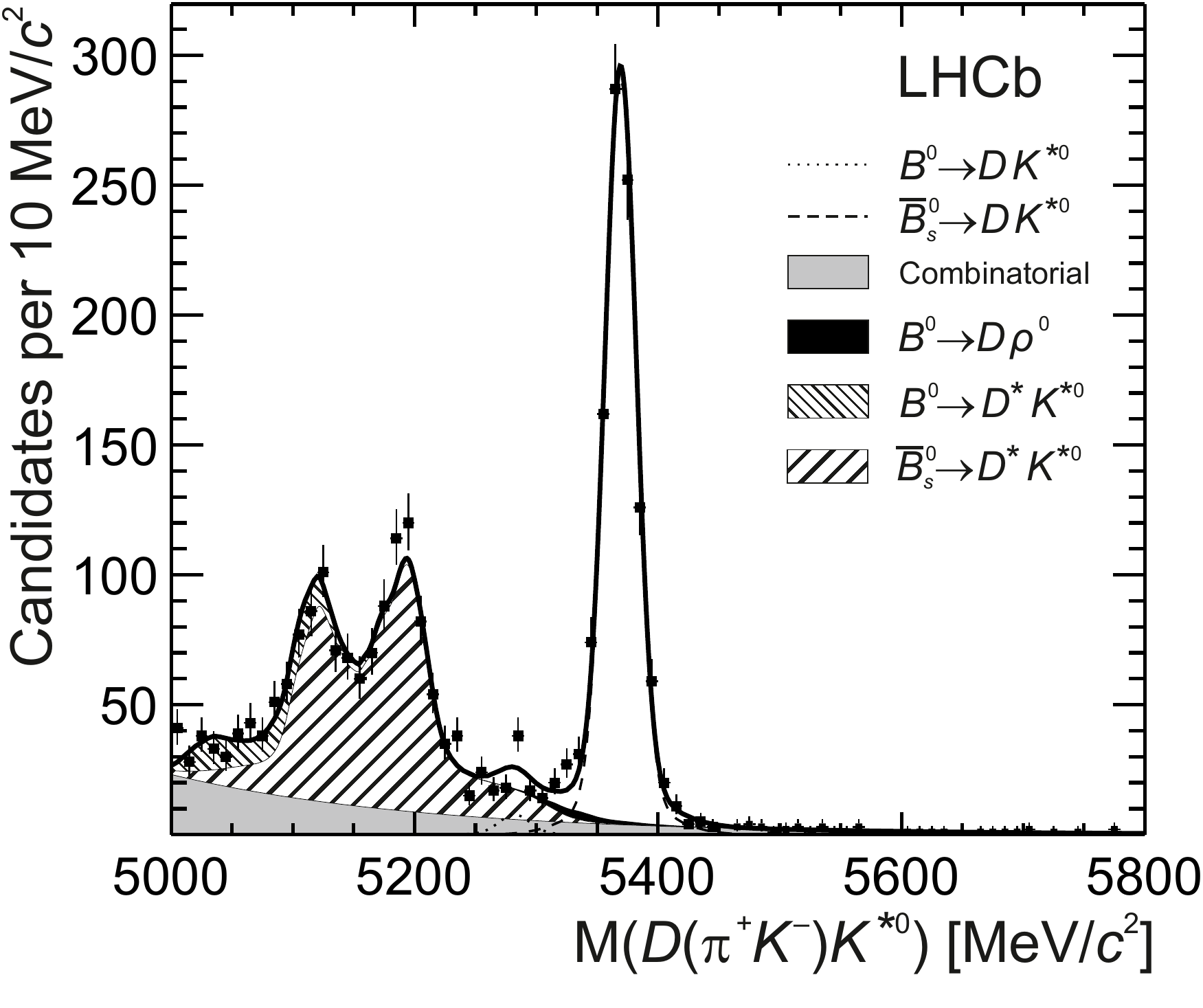}

\includegraphics[width=0.49\textwidth]{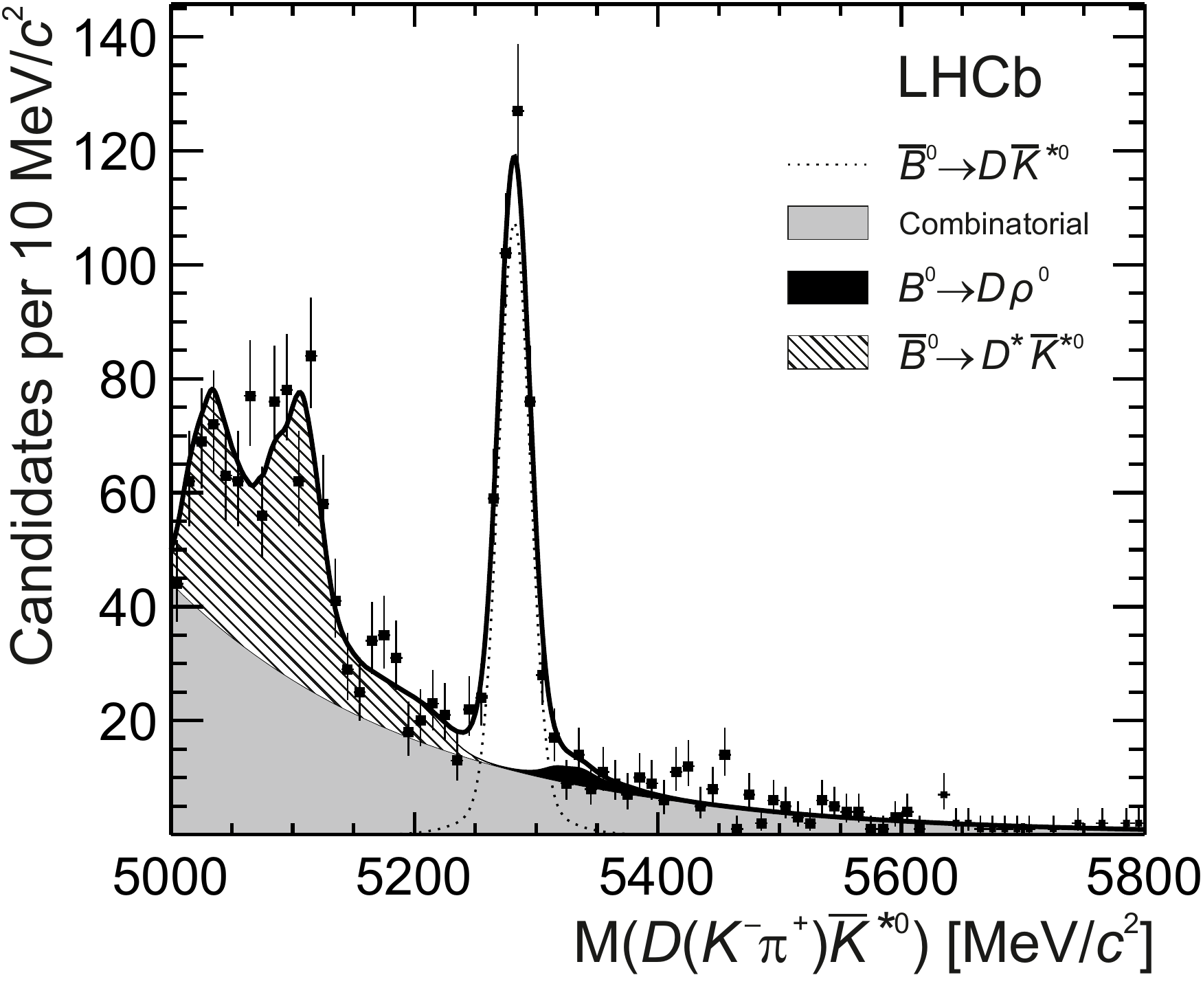}
\includegraphics[width=0.49\textwidth]{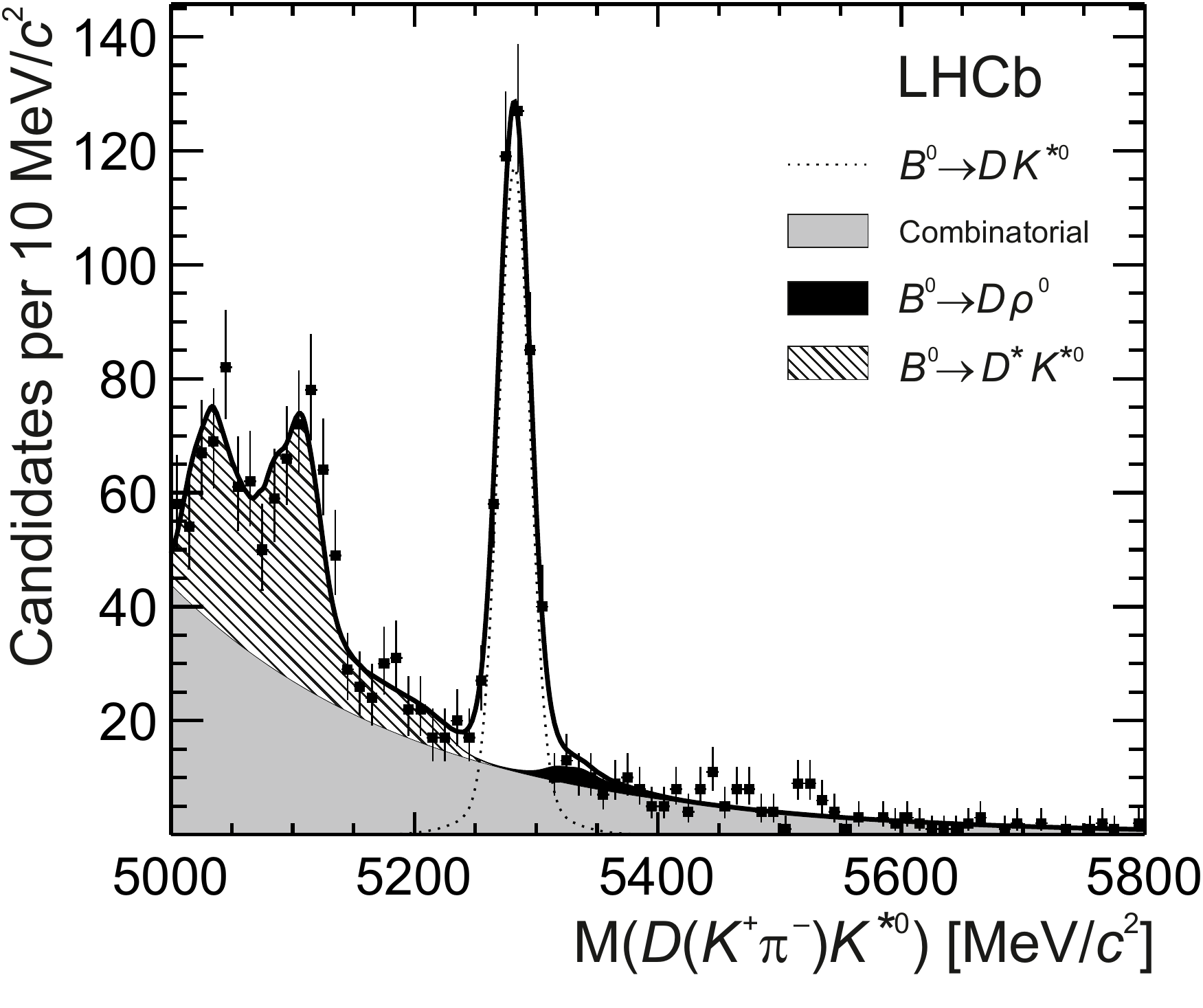}
\caption{\small Distributions of (top left)  $\D(\pim \Kp) \Kstarzb$, (top right) $\D(\pip \Km) \Kstarz$, (bottom left) $\D(\Km\pip) \Kstarzb$  and (bottom right)  $\D(\Kp\pim) \Kstarz$ invariant mass. 
The data (black points) and the fitted invariant mass model (thick solid line) are shown. The PDFs corresponding to the different species are indicated in the legend: the $\Bz$ signal, the $\Bs$ signal, combinatorial background, $\Bz\to\D \rho^0$ background, partially reconstructed $\Bs\to \Dstar\Kstarzb$ and $\Bz\to \Dstar \Kstarz$ backgrounds.}
\label{fig:plotKPiOS}
\end{figure}

\begin{figure}[!b]
\centering
\includegraphics[width=0.49\textwidth]{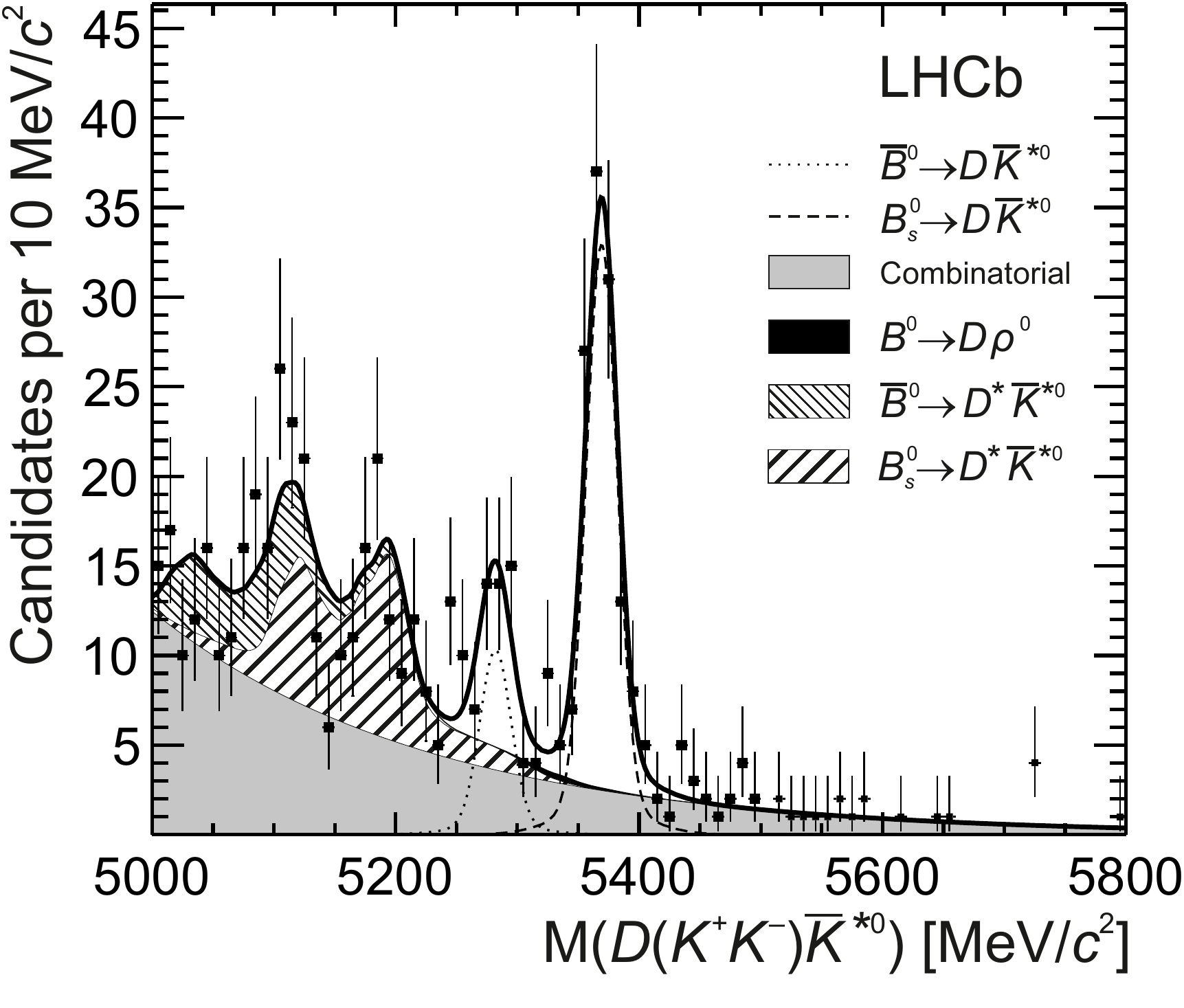}
\includegraphics[width=0.49\textwidth]{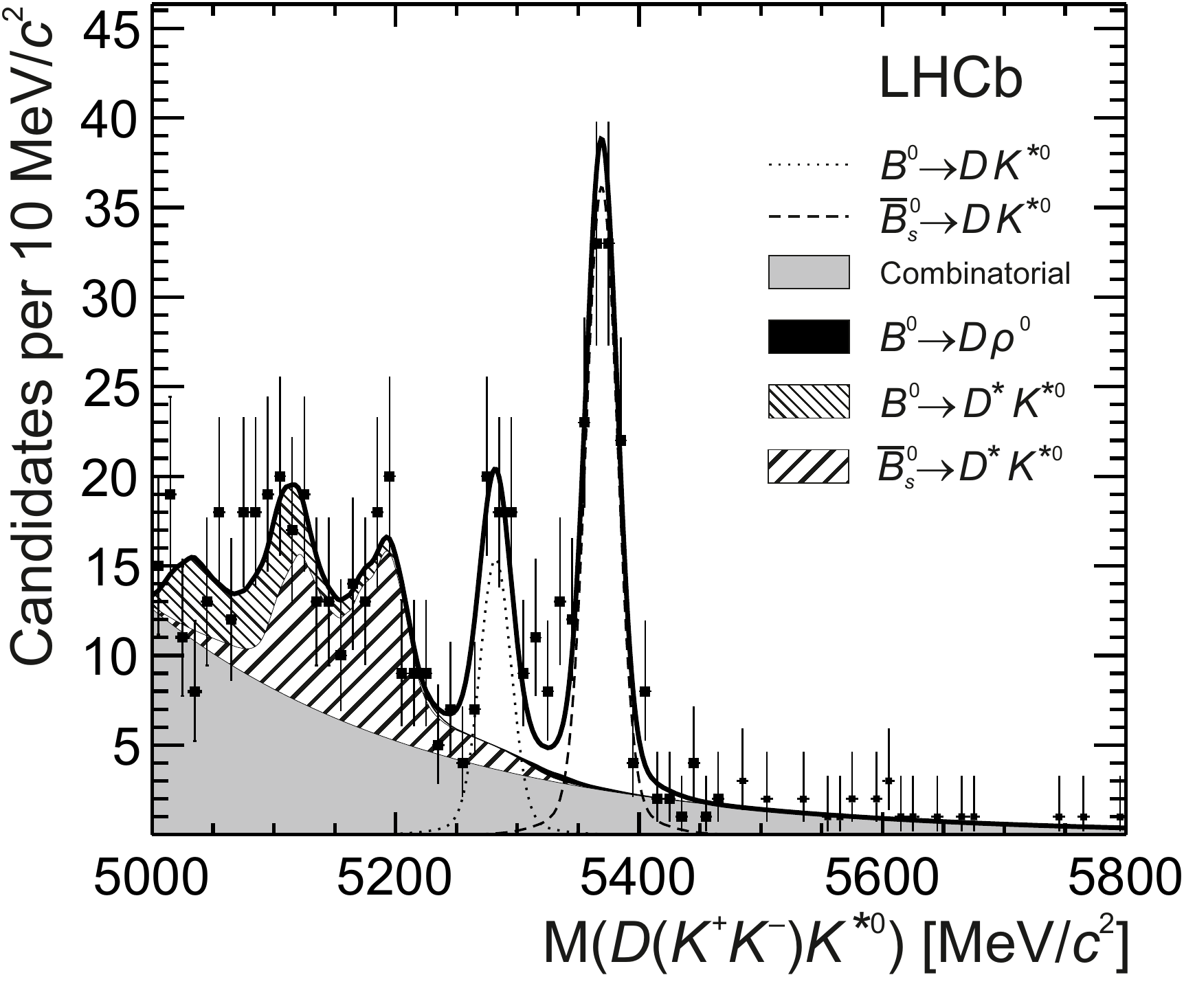}

\includegraphics[width=0.49\textwidth]{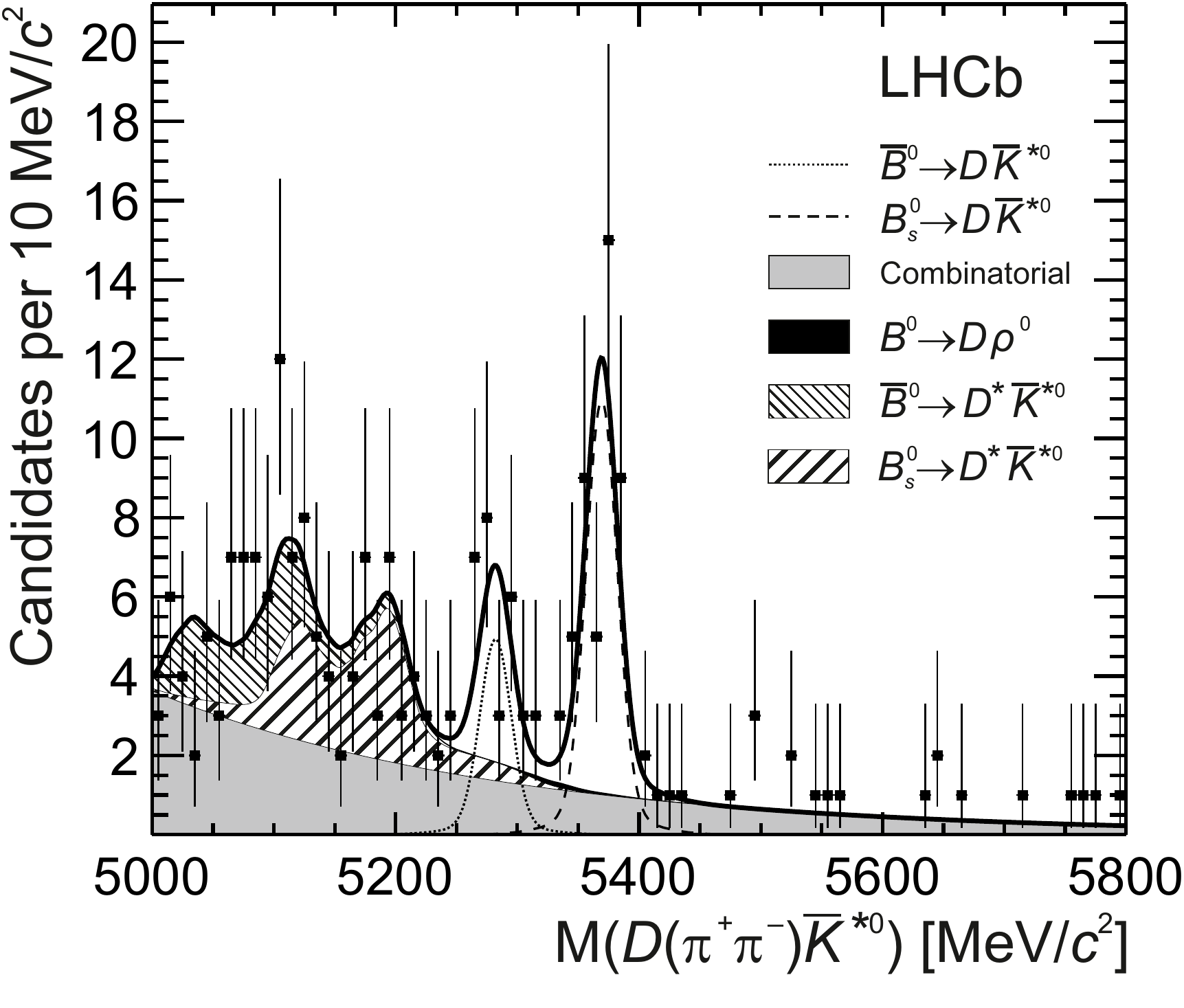}
\includegraphics[width=0.49\textwidth]{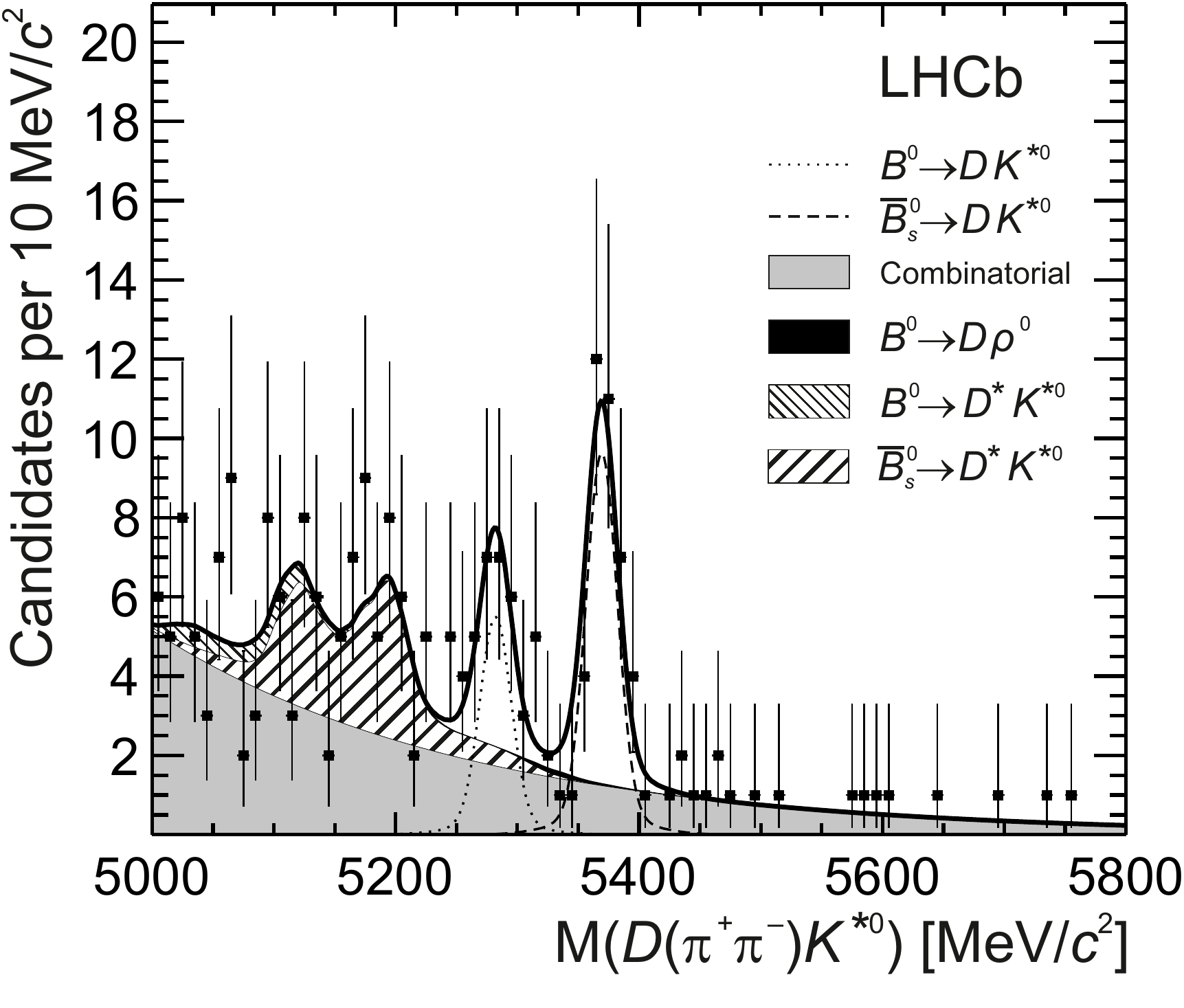}
\caption{\small Distributions of (top left) $\D(\Kp\Km) \Kstarzb$, (top right) $\D(\Kp\Km) \Kstarz$, (bottom left) $\D(\pip\pim) \Kstarzb$ and (bottom right) $\D(\pip\pim) \Kstarz$  invariant mass. 
The data (black points) and the fitted invariant mass model (thick solid line) are shown. The PDFs corresponding to the different species are indicated in the legend: the $\Bz$ signal, 
the $\Bs$ signal, combinatorial background, $\Bz\to\D \rho^0$ background and partially reconstructed $\Bs\to \Dstar\Kstarzb$ and $\Bz\to \Dstar \Kstarz$ backgrounds.}
\label{fig:plotPiPi}
\end{figure}

The non-parametric functions used to model all the specific backgrounds are smeared to take into account the different mass resolutions observed in data 
and simulation.
The invariant mass distributions together with the function resulting from the fit are shown in Figs.~\ref{fig:plotKPiOS} and \ref{fig:plotPiPi}. The numbers of 
signal events in each category are summarised in Table~\ref{tab:fitresult}.

\renewcommand{\arraystretch}{1.4}
\begin{table}[!t]
\tabcolsep 3mm
\begin{center}
\caption{\small \label{tab:fitresult}Yields of signal candidates with their statistical uncertainties.}
\begin{tabular}{@{}llll@{}}
\toprule
Channel & Signal yield & Channel & Signal yield \\
\midrule

$\Bzb\to \D(\pim\Kp) \Kstarzb$ & $\phantom{1}24\pm 12$ &
$\Bz\to \D(\pip\Km) \Kstarz$ & $\phantom{1}26\pm 12$  \\

$\Bzb\to \D(\Km\pip) \Kstarzb$ & $370  \pm 22$ &
$\Bz\to \D(\Kp\pim) \Kstarz$ & $405  \pm 23$  \\

$\Bzb\to \D(\Kp\Km) \Kstarzb$ & $\phantom{1}36 \pm 9$ & 
$\Bz\to \D(\Kp\Km) \Kstarz$ & $\phantom{1}53 \pm 10$  \\

$\Bzb\to \D(\pip\pim) \Kstarzb$ & $\phantom{1}18 \pm 6$ &
$\Bz\to \D(\pip\pim) \Kstarz$ & $\phantom{1}21 \pm 7$  \\

\midrule

$\Bs\to \D(\pim\Kp) \Kstarzb$ & $933 \pm 33$ & 
$\Bsb\to \D(\pip\Km) \Kstarz$ & $993 \pm 34$ \\ 

$\Bs\to \D(\Kp\Km) \Kstarzb$ & $115 \pm 12$ & 
$\Bsb\to \D(\Kp\Km) \Kstarz$ & $125 \pm 13$ \\

$\Bs\to \D(\pip\pim) \Kstarzb$ & $\phantom{1}39 \pm 7$ & 
$\Bsb\to \D(\pip\pim) \Kstarz$ & $\phantom{1}35 \pm 7$ \\

\bottomrule
\end{tabular}
\end{center}
\end{table}
\renewcommand{\arraystretch}{1.}

\section{Systematic uncertainties}
\label{sec:systematics}

The signal yields determined from the invariant mass fit are corrected in order to evaluate the asymmetries and ratios described 
in~\eqref{eq:R_CPplus}-\eqref{eq:A_s_piK}. These corrections account for selection efficiency and detection asymmetry, 
$\Bb$-$\B$ production asymmetry and its dilution due to
mixing, misidentification of \D meson decays, \Dz decay branching fractions, hadronisation fractions and biases introduced by the fit model. The uncertainties in 
these corrections cause systematic uncertainties in the results. Systematic uncertainties are also introduced by the uncertainties in the various constraints on the 
invariant mass model. The systematic uncertainties incurred from all sources are obtained combining in quadrature the individual uncertainties and
are summarised in Table~\ref{tab:Final_Systematics_1}.

\subsection{Efficiencies}
\label{subsec:efficiency_corrections}

Separate corrections are applied to account for differing trigger and PID efficiencies. 
These efficiencies are obtained from real data by means of low-background calibration samples of kaons and 
pions from $\Dstarpm\to\D(\Kmp\pipm)\pipm$ decays~\cite{LHCb-PUB-2011-026}. They are evaluated separately for \Bb and \B modes to account for 
detection asymmetries. 
The relative trigger and PID efficiencies differ from unity by 1\% and 5\%, respectively, and their uncertainties result in the systematic uncertainties given in 
Table~\ref{tab:Final_Systematics_1}. 

Another correction is applied to account for the differences in the kinematic selection requirements of the different decay modes.
The efficiencies are evaluated from simulated data and they are assumed to be equal for the $\Bz\to\D(\Kp\pim)\Kstarz$ and $\Bz\to\D(\pip\Km)\Kstarz$  decays.
They differ between decay modes by 8\% at maximum. The uncertainties on these efficiencies affect the measured observables as shown in Table~\ref{tab:Final_Systematics_1}.
It is noted that the ${\cal R}_d^\pm$ observables have no systematic uncertainty from 
selection efficiency. This is because they are separated by \B meson flavour and have the same \D meson final state; therefore all efficiencies are assumed to 
cancel.

Because of the different \Bz and \Bs lifetimes, the ratio of efficiencies for $\Bz\to\D(h^+h^{\prime-})\Kstarz$ to $\Bsb\to\D(h^{\prime+}h^{-})\Kstarz$ is different 
from one. This ratio is assumed to be equal between all the \D meson final states and is calculated using the $\Bz\to\D(\Kp\pim)\Kstarz$ and 
$\Bsb\to\D(\pip\Km)\Kstarz$ decay modes, assuming that the lifetime difference effects factorise from the other selection effects. 
The difference in \Bz and \Bs selection efficiencies arises from the use of variables sensitive to the decay topology in the BDT and
is equal to 3\%.
The systematic uncertainty from this source is labelled ``Lifetime difference'' in Table~\ref{tab:Final_Systematics_1}.
The only observables affected by the systematic uncertainty due to lifetime difference are the ${\cal R}_{ds}^{hh}$ observables, since only these 
involve both \Bz and \Bs partial widths. 

\subsection{Production asymmetry}
\label{subsec:prod_asymm_correction}

The difference between \Bz and \Bzb, or \Bs and \Bsb, production rates in $pp$ collisions is accounted for by applying a correction factor
$a_P = (1-\alpha A_P)/(1+\alpha A_P)$ to the \Bzb and \Bsb signal yields, where 
\begin{equation}
A_P\equiv \frac{\sigma(\Bb) - \sigma(\B)}{\sigma(\Bb) + \sigma(\B)}
\end{equation}
is the raw production asymmetry of 
the \Bz or \Bs mesons in question. In the case of \Bz mesons, 
$A_P$ has been measured, using $\Bz\to \jpsi\Kstarz$ decays, to be $A_P=0.010\pm 0.013$~\cite{LHCb-PAPER-2011-029}. The effect of the raw production 
asymmetry on the number of observed \Bzb or \Bz decays becomes less pronounced for larger decay times due to mixing. It is also affected by the selection 
efficiency as a function of the decay time, $\epsilon(\Bz\to\D \Kstarz,t)$. A factor, $\alpha$, accounts for this dilution and  is given for \Bz mesons by
\begin{equation}
\label{eq:alpha_B0}
\alpha=\frac{\int_0^{+\infty}e^{-t/{\tau_\Bz}}\cos(\Delta m_d t)\epsilon(\Bz\to\D \Kstarz,t)\,{\rm d}t}{\int_0^{+\infty}e^{-t/{\tau_\Bz}}\epsilon(\Bz\to\D \Kstarz,t)\,{\rm d}t},
\end{equation}
where $\Delta m_d$ is the \Bz-\Bzb oscillation frequency and $\tau_\Bz$ is the \Bz lifetime.

The factor $\alpha$ is evaluated separately for each $\Bz\to \D(\Kpm\pimp)\Kstarz$, $\Bz\to \D(\Kp\Km)\Kstarz$ and $\Bz\to \D(\pip\pim)\Kstarz$ decays since it 
is dependent on the separately optimised selection requirements. The resulting values of $\alpha$ are $0.362 \pm 0.014$, $0.391 \pm 0.014$ and 
$0.398 \pm 0.014$, respectively. These figures are computed using fully simulated events and data-driven PID efficiencies from calibration samples. 
The uncertainty 
on $a_P$ is propagated to the measured observables to estimate the systematic uncertainty from the production asymmetry and mixing. 
Owing to the 
large \Bs oscillation frequency, a potential production asymmetry of \Bs mesons does not significantly affect the measurements presented here and is 
neglected.

\subsection{Misidentification of \D meson decays}
\label{subsec:D0_doubleMisID}

Favoured $\Bz\to\D(\Kp\pim)\Kstarz$ decays are misidentified as suppressed $\Bz\to\D(\pip\Km)\Kstarz$ decays at a  small but non-negligible rate.  
The fraction of signal $\Bz\to\D(\Kp\pim)\Kstarz$ decays reconstructed as signal $\Bz\to\D(\pip\Km)\Kstarz$ decays is estimated from the simulation
to be less than 1\% after
applying the veto described in Sect.~\ref{sec:selection}. 
However, the best-fit values of the numbers of $\Bz\to\D(\pip\Km)\Kstarz$ decays are corrected to take this into 
account. The uncertainty in this correction causes a systematic uncertainty in the ${\cal R}_d^\pm$ observables given in Table~\ref{tab:Final_Systematics_1}
as misID.

\subsection{Other corrections}
\label{subsec:other_corrections}

Two ratios of \Dz meson decay branching fractions (BF) are needed to compute the final results because of the approximation made between ${\cal R}_{\CP+}$ and ${\cal R}_d^{hh}$ in Eq.~\eqref{eq:R_d_hh}. These are taken from Ref.~\cite{PDG2012}, the results of which imply that the ratio of $\BR\left(\Dz\to\Km\pip\right)$ to $\BR\left(\Dz\to\Kp\Km\right)$ is $9.80 \pm 0.24$ and the ratio of $\BR\left(\Dz\to\Km\pip\right)$ to $\BR\left(\Dz\to\pip\pim\right)$ is $27.7 \pm 0.6$.

The fraction of \bquark quarks that hadronize into $\Bz$ and $\Bs$ mesons in $pp$ collisions, $f_d$ and $f_s$, respectively,  
has an effect on the number of $\Bz$ and $\Bs$ mesons produced in \lhcb. 
Since the ${\cal R}_{ds}^{hh}$ observables are ratios of $\Bd$ and $\Bs$ decay partial widths, they are corrected 
with the hadronisation fraction ratio $f_s/f_d=0.267\pm0.021$~\cite{LHCb-PAPER-2011-018}. 
The ${\cal R}_{ds}^{hh}$ observables also contain a factor of $\tau_{\Bs}/\tau_{\Bz}$, which arises because of the lifetimes, $\tau$, of the \Bz and \Bs mesons. 
This is taken from Ref.~\cite{PDG2012}, the results of which imply that $\tau_{\Bs}/\tau_{\Bz}=0.99 \pm 0.01$.

\renewcommand{\arraystretch}{1.3}
\begin{table}[!tb]
\tabcolsep 2mm
\begin{center}
\caption{\small Uncertainties in the observables. All model-related systematic 
uncertainties are added in quadrature and the result is shown as one source of systematic uncertainty. The presence of `--' indicates that the source of 
uncertainty does not affect the observable.}
\label{tab:Final_Systematics_1}
\begin{small}
\begin{tabular}{@{}l@{\,}c@{\,\,\,}c@{\,\,\,}c@{\,\,\,}c@{\,\,\,}c@{\,\,\,}c@{\,\,\,}c@{\,\,\,}c@{\,\,\,}c@{\,\,\,}c@{\,\,\,}c@{\,\,\,}c@{}}
\toprule
Source & \multicolumn{12}{c}{Observable} \\ 
& ${\cal{A}}^{KK}_{d}$ & ${\cal{A}}^{\pi \pi}_{d}$ & ${\cal{R}}^{KK}_{d}$ & ${\cal{R}}^{\pi \pi}_{d}$ & ${\cal{R}}_d^{+}$ & ${\cal{R}}_d^{-}$  & ${\cal{R}}^{KK}_{ds}$  & ${\cal{R}}^{\pi \pi}_{ds}$ & ${\cal{A}}^{KK}_{s}$ & ${\cal{A}}^{\pi \pi}_{s}$ & ${\cal{A}}_{d}^{K\pi}$   & ${\cal{A}}^{\pi K}_{s}$   \\ 
\midrule 
Trigger efficiency & 0.011 & 0.011 & 0.015 & 0.019 & -- & --  & 0.000  & 0.000  & 0.011 & 0.011 & 0.012   & 0.012   \\ 
PID efficiency & 0.005  & 0.005 & 0.010 & 0.012 & -- & --  & 0.000  & 0.000 & 0.005 & 0.005 & 0.005  & 0.005    \\ 
Selection efficiency & 0.014 & 0.014 & 0.029 & 0.037 & -- & --  & 0.000 & 0.000 & 0.015 & 0.014 & 0.014    & 0.014     \\ 
Lifetime difference & --  & -- & -- & -- & -- & --  & 0.002 & 0.003 & -- & --    & --  & --  \\ 
Prod. asymmetry & 0.005 & 0.005 & 0.001 &  0.000 & -- & --  & 0.000 & 0.001 & -- & --  & 0.005    & --   \\ 
$\D\to K\pi$ misID & -- & --   & --  & -- & 0.000 & 0.001 & -- & -- & -- & --   & --    & --  \\ 
$D^0$ decay BFs & -- & -- & 0.025 & 0.028  & -- & -- & --  & --  & --  & --  & --  & --   \\ 
$f_s/f_d$ & -- & --  & -- & --  & -- & --  & 0.008 & 0.012 & -- & --    & -- & -- \\ 
$\tau_s/\tau_d$ & -- & -- & -- & --  & -- & --  & 0.001 & 0.001  & -- & --  & --  & --   \\ 
Model-related & 0.004 & 0.001 & 0.011 & 0.012 & 0.010 & 0.011  & 0.002 & 0.001 & 0.004 & 0.002 & 0.001   & 0.000    \\ 
\midrule 
Total systematic & 0.020 & 0.019 & 0.044 & 0.053 & 0.010 & 0.011  & 0.009 & 0.012 & 0.020 & 0.019 & 0.020    & 0.019    \\ 
\midrule 
Statistical & 0.144 & 0.217 & 0.159 & 0.268 & 0.028 & 0.031  & 0.017 & 0.038 & 0.073 & 0.131 & 0.041  & 0.025   \\ 
\bottomrule
\end{tabular}
\end{small}
\end{center}
\end{table}
\renewcommand{\arraystretch}{1.}

\subsection{Model-related systematic uncertainty}
\label{subsec:bias_correction}

The \B meson invariant mass model is validated with an ensemble of simulated pseudoexperiments. The results of these pseudoexperiments 
show small biases, of the order 
of 1\% of the statistical uncertainty, in the best-fit values of the signal yields, as determined by the invariant mass fit. 
The affected signal yields are corrected for these biases before computing the observables. The statistical uncertainty on the bias due to the limited number of 
pseudoexperiments causes systematic uncertainty in the observables. 

Systematic uncertainties due to the effects of the constraints made when constructing the invariant mass fit model are also evaluated with 
pseudoexperiments. The constraints considered are 
\begin{enumerate}
\item The values fixed from simulation of the core fraction and the ratio between the widths of the two Gaussian functions used as signal PDF;
\item The difference in mass of the \Bz and \Bs mesons from Ref.~\cite{PDG2012};
\item The branching ratios from Ref.~\cite{PDG2012} and selection efficiencies from simulation used to constrain the relative normalisations of the background PDFs.
\end{enumerate}

Each fixed parameter of the model has an associated uncertainty. To evaluate this, the invariant mass model is altered such that a 
particular fixed parameter is varied by 
its uncertainty and data sets generated with the default model are fitted with this altered value. The variations in the best-fit values of the signal yields observed 
when changing the model are used to assign a systematic uncertainty on the signal yields. This process is repeated for each fixed parameter and the systematic 
uncertainties in the signal yields are propagated to the observables.
All model-related systematic uncertainties are added in quadrature and this figure is given in Table~\ref{tab:Final_Systematics_1}.

\section{Results}

The results are
\begin{center}%
\tabcolsep 5mm%
\renewcommand{\arraystretch}{1.7}
\begin{tabular}{l@{}l@{}ll@{}l@{}l}%
${\cal{A}}^{KK}_{d}$ & $\,=\,\,$ & $-0.20  \, \pm 0.15 \, \pm 0.02$, &
${\cal{A}}^{\pi \pi}_{d}$ & $\, =\,\, $ & $-0.09  \, \pm 0.22 \, \pm 0.02$, \\
${\cal{R}}^{KK}_{d}$ & $\, = \,\,$ & $ \phantom{-}1.05  \, ^{+0.17}_{-0.15} \, \pm 0.04$, &
${\cal{R}}^{\pi \pi}_{d}$ & $\, = \,\,$ & $\phantom{-}1.21  \, ^{+0.28}_{-0.25} \, \pm 0.05$,  \\
${\cal{R}}^{+}_{d}$ & $\, = \,\,$ & $\phantom{-}0.06  \, \pm 0.03 \, \pm 0.01$, & 
${\cal{R}}^{-}_{d}$ & $\, = \,\,$ & $\phantom{-}0.06  \, \pm 0.03 \, \pm 0.01$, \\
${\cal{R}}^{KK}_{ds}$ &  $\,=\,\,$ & $ \phantom{-}0.10  \, \pm 0.02 \, \pm 0.01$, & 
${\cal{R}}^{\pi \pi}_{ds}$ & $\, =\, \,$ & $\phantom{-}0.15  \, \pm 0.04 \, \pm 0.01$, \\ 
${\cal{A}}^{KK}_{s}$ & $ \, = \,\,$ & $ -0.04 \pm 0.07 \, \pm 0.02$,  & 
 ${\cal{A}}^{\pi \pi}_{s}$ & $\, = \,\,$ & $ \phantom{-}0.06  \, \pm 0.13 \, \pm 0.02$, \\ 
${\cal{A}}_{d}^{K\pi}$ &  $\,=\,\,$  & $-0.03  \, \pm 0.04 \, \pm 0.02$,  & 
${\cal{A}}^{\pi K}_{s} $ & $\, = \,\,$ & $-0.01  \, \pm 0.03 \, \pm 0.02$, 
\end{tabular}
\renewcommand{\arraystretch}{1.0}
\end{center}
where the first uncertainties are statistical and the second systematic~\cite{correlations}. The significances of the combined
\Bz and \Bzb signals for the $\Bz\to\D(\pip\Km)\Kstarz$, $\Bz\to\D(\Kp\Km)\Kstarz$ and $\Bz\to\D(\pip\pim)\Kstarz$ decay modes are
$2.9\sigma$, $8.6\sigma$ and $5.8\sigma$, respectively, including systematic uncertainties. The statistical significances, 
expressed in terms of number of standard deviations ($\sigma$), are
computed from $\sqrt{2\ln(L_{\rm sig}/L_0)}$ where $L_{\rm sig}$ and $L_0$ are the likelihoods from the nominal mass fit described
in Sect.~\ref{sec:fit} and from the same fit omitting the signal component, respectively. The likelihoods are convolved
with a Gaussian function of width equal to the systematic uncertainties on the fit model in order to compute the total significances.
No significant \CP violation effect is observed. 

The constraints from the measurements pertaining
to \Bz mesons on the angle $\gamma$ of the unitarity triangle and the hadronic parameters $r_B$ and $\delta_B$ are presented in 
Sect.~\ref{sec:interpretation}.
With more data, improved measurements of the quantities related to $\Bs \to \D \Kstarzb$
decays will also contribute to the sensitivity but are not used here.

\section{Implication on the value of $\boldsymbol{r_B}$}
\label{sec:interpretation}

The sensitivity of these results to the CKM phase $\gamma$ is investigated by employing a frequentist method described in 
Ref.~\cite{LHCb-PAPER-2013-020} to scan the $(\gamma,r_B,\delta_B)$ parameter 
space and calculate the \chisq probability at each point, given the measurements of the observables
and using~\eqref{eq:R_d_hh_1}-\eqref{eq:R_d_plusminus_1}. The statistical and systematic uncertainties are combined in 
quadrature and their correlations are accounted for. 
In principle, the coherence factor $\kappa$ can also be extracted together with $\gamma$, $r_B$ and $\delta_B$ but
the uncertainties of the measurements are too large with the current data sample size to constrain all parameters together. 
A value of $\kappa=0.95\pm0.03$ is used instead. 
This value is determined from a toy simulation study of a realistic model for the resonance content of $\Bz\to \D \Kp \pim$ decays, similar to the method
used in Ref.~\cite{Pruvot:2007yd}. This model describes the decay amplitude in the analysis phase space as a superposition of a non-resonant component 
and amplitudes corresponding to the intermediate $\Kstar(892)^0$, $\Kstar(1410)^0$, $K_0^*(1430)^0$, $K_2^*(1430)^0$,
$\Kstar(1680)^0$, $D_0^*(2410)^-$, $D_2^*(2460)^-$ and $D_{s2}(2573)^+$ resonances. The relative fractions and phases between these
components are generated randomly according to their known values and uncertainties~\cite{PDG2012} when they have been observed or within conservatively
large ranges when they have not been measured. 
The analysis selection effects are taken into account, and the main requirements affecting the value of $\kappa$ are
the $K^*(892)^0$ mass selection of 50\mevcc around the known mass and the selection on $|\cos\theta^*|$ being larger than 0.4. 
The $\Dz\to\Kpm\pimp$ amplitude ratio $r_D$ and strong phase difference $\delta_D$ are taken from the Heavy Flavour Averaging
Group~\cite{HFAG}.

\begin{figure}[!h]
\centering
\includegraphics[width=0.6\textwidth]{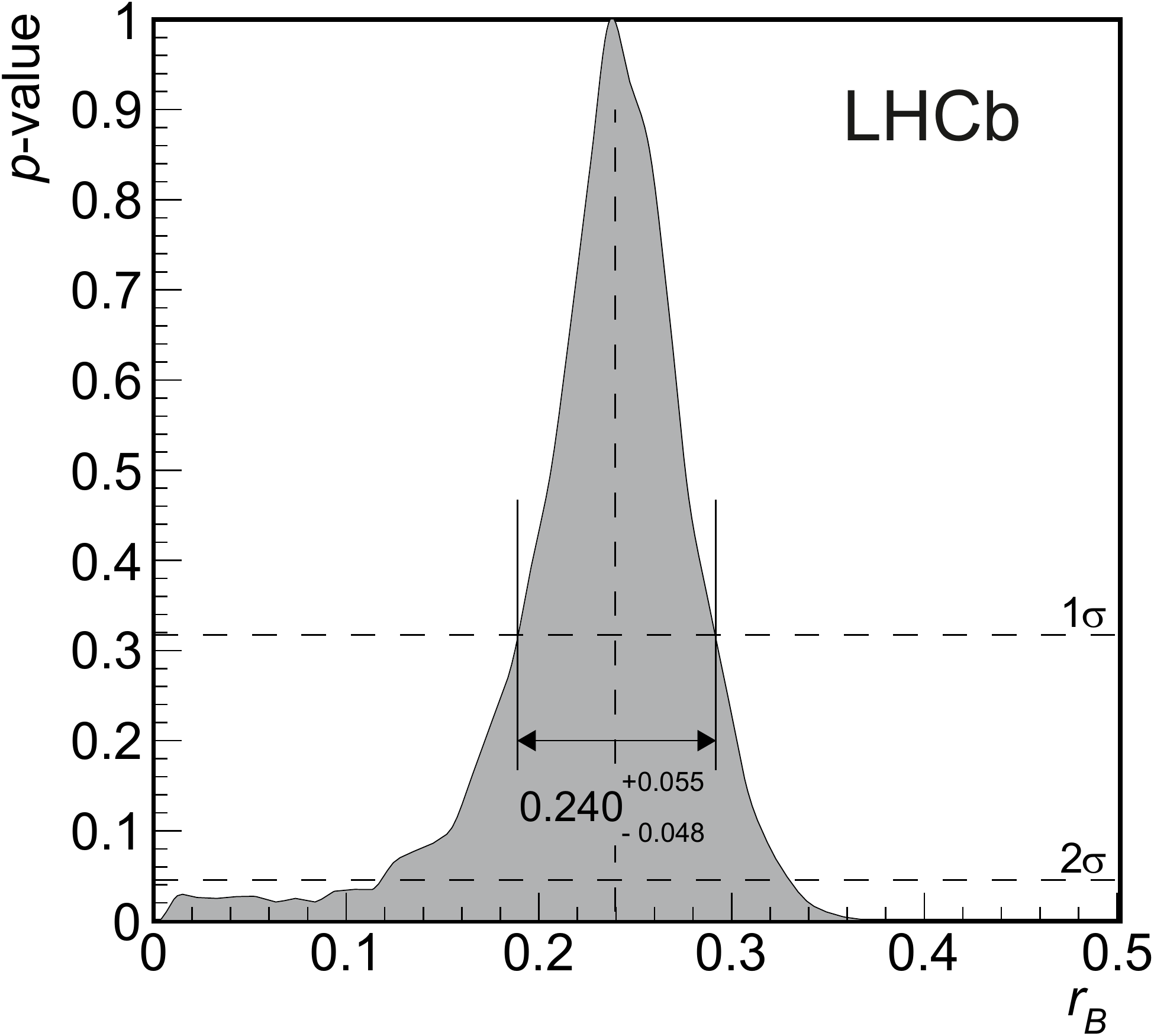}
\caption{$p$-value as a function of $r_{\B}$, for $\kappa=0.95\pm0.03$. 
The horizontal dashed lines represent the $1\sigma$ and $2\sigma$ confidence levels and the vertical dotted line represents the obtained central value.}
\label{fig:1D_CLplot_rB}
\end{figure}

A one-dimensional projection of the $p$-value, or $1-{\rm CL}$, is given in Fig.~\ref{fig:1D_CLplot_rB}, which shows that $r_B$ is
\begin{equation}
r_B=0.240\,^{+0.055}_{-0.048} \nonumber
\end{equation}
at a confidence level of 68.3\% and is different from 0 with a significance of $2.7\sigma$. 
The $p$-value at each point of $r_B$ is computed with simulated pseudoexperiments following a Feldman-Cousins method, where the nuisance 
parameters are kept at their best-fit values obtained at each point of $r_B$.

\begin{figure}[!h]
\centering
\includegraphics[width=0.49\textwidth]{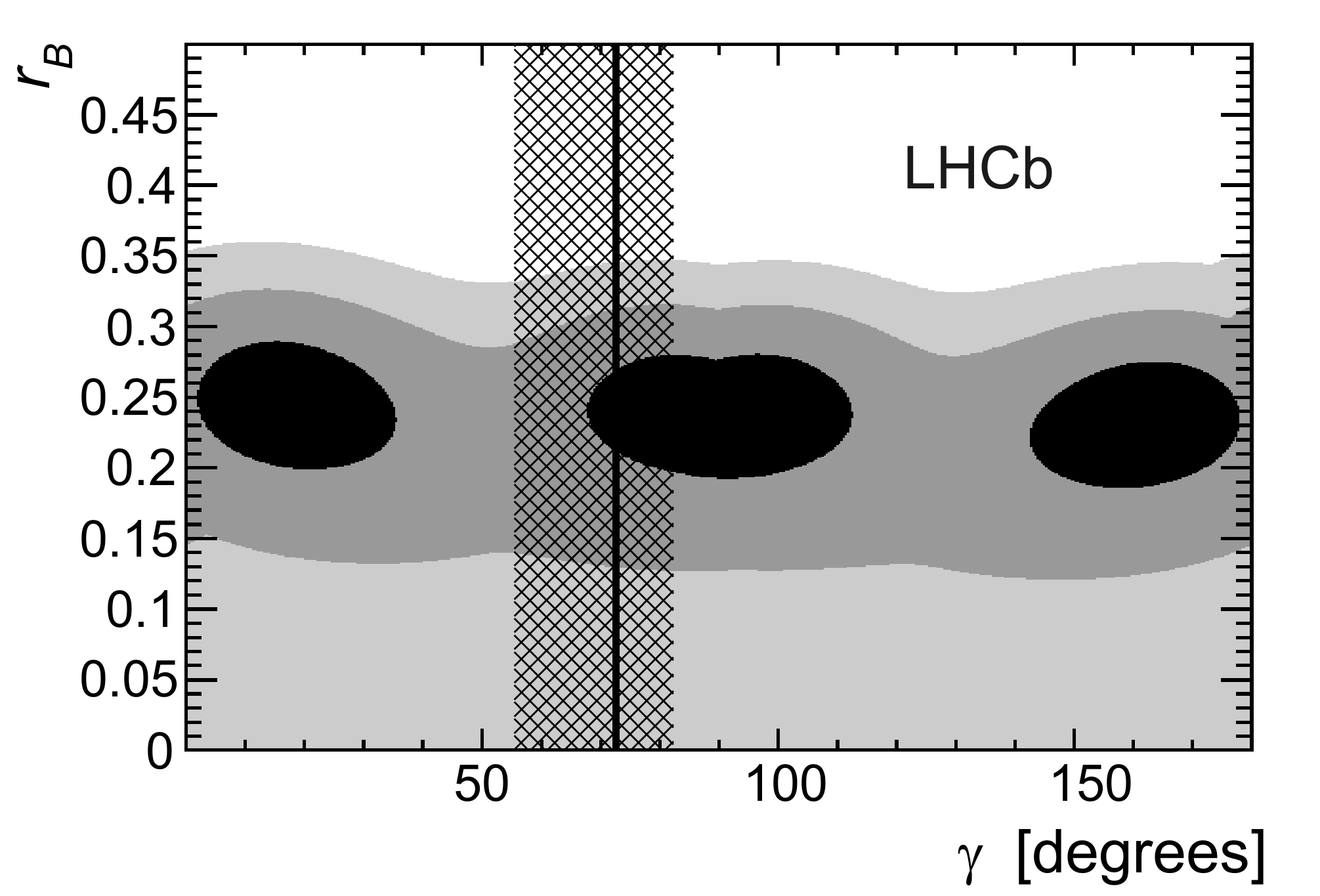}
\includegraphics[width=0.49\textwidth]{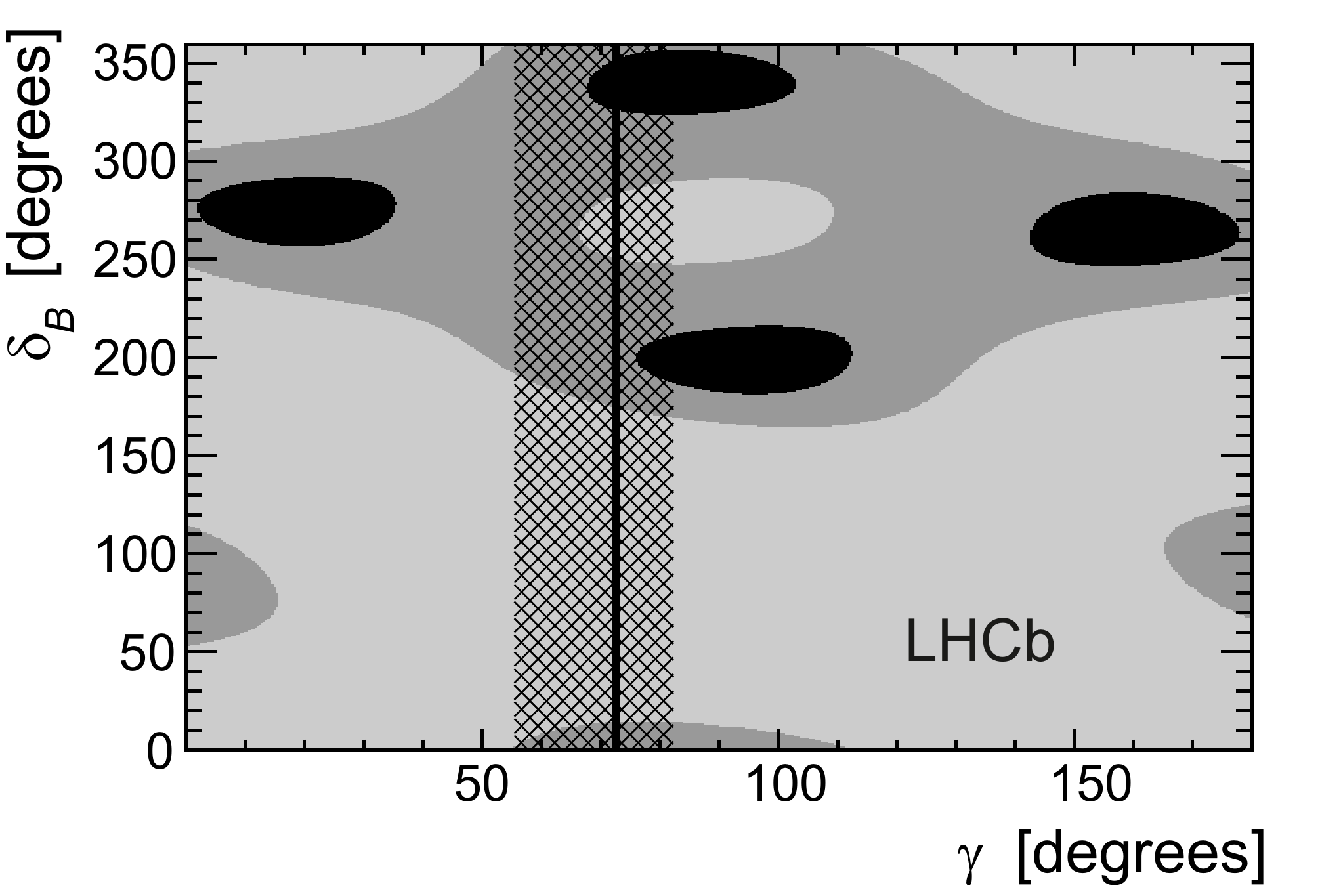}
\caption{Two-dimensional projections of the $p$-value  in 
$(r_B,\delta_B,\g)$ parameter space onto (left) $r_{\B}$ and $\gamma$ and (right) $\delta_{\B}$ and $\gamma$, for $\kappa=0.95\pm0.03$. The contours are the $n\sigma$ 
profile likelihood contours, where $\Delta\chi^2 = n^2$ with $n = 1$ (black), 2 (medium grey), and 3 (light gray), 
corresponding to 39.4\%, 86.5\% and 98.9\% confidence level, respectively. The vertical line and hashed band represent the best-fit value of \g and the 
68.3\% confidence level interval by Ref.~\cite{LHCb-PAPER-2013-020}.}
\label{fig:gammadini}
\end{figure}

Two-dimensional projections of the $p$-value from the profile likelihood are shown
in Fig.~\ref{fig:gammadini}. The  \lhcb average value for $\gamma$, 
extracted from a combination of $\Bpm\to\D\Kpm$ and $\Bpm\to\D\pipm$ analyses~\cite{LHCb-PAPER-2013-020}, is shown with its 68.3\% confidence level interval.
The precision of the current results does not allow a significant measurement of $\gamma$ from $\Bz\to\D\Kstarz$ decays alone, but these measurements 
could nonetheless be used in a global fit.

\section{Conclusions}
\label{sec:conclusions}

The parameters of the $\Bz\to\D\Kstarz$ decay, which are sensitive to the CKM angle $\gamma$, have been measured with a sample 
of 3.0\invfb of LHC $pp$ collision data collected by the LHCb detector. The results include the first measurements
of \CP asymmetries in \Bz and \Bsb to $\D \Kstarz$ decays with the neutral \D meson decaying into the $\pip\pim$ final state. 
The results related to the $\Kp\Km$ final state of the \D meson, ${\cal A}_d^{KK}$ and ${\cal R}_d^{KK}$, are
in agreement with and more precise than those from a previous analysis of \lhcb data~\cite{LHCb-PAPER-2012-042},
and supersede them.
The measurements of ${\cal R}_d^+$ and ${\cal R}_d^-$ presented here are the first
obtained separately for \Bz and \Bzb mesons. 
They are consistent with the measurement of the flavour-averaged ratio 
\begin{equation}
\frac{\Gamma(\Bzb\to\D(\pim\Kp)\Kstarzb)+\Gamma(\Bz\to\D(\pip\Km)\Kstarz)}{\Gamma(\Bzb\to\D(\Km\pip)\Kstarzb)+\Gamma(\Bz\to\D(\Kp\pim)\Kstarz)}
\end{equation}
by the \belle collaboration~\cite{BELLE_DKstar} using the same $\Kstarz$ invariant mass range.

From the measurements presented in this article, we measure the value of $r_{\B}(\D\Kstarz)$, the ratio of the amplitudes of the decay $\Bz\to\D\Kp\pim$
with a $\bquark\to\uquark$ or a $\bquark\to\cquark$ transition,
in a $K\pi$ mass region of $\pm50\mevcc$ around the $K^*(892)^0$ mass, and for an absolute value of the cosine of the \Kstarz helicity angle larger than 0.4.
It is found to be equal to $0.240_{-0.048}^{+0.055}$ at a confidence level of 68.3\%. 
This is the first measurement of this parameter with \lhcb data and is more accurate than the previous measurement
made by the \babar collaboration~\cite{BABAR_DKstar}, in a comparable region of phase space.


\section*{Acknowledgements}

\noindent  We express our gratitude to our colleagues in the CERN
accelerator departments for the excellent performance of the LHC. We
thank the technical and administrative staff at the LHCb
institutes. We acknowledge support from CERN and from the national
agencies: CAPES, CNPq, FAPERJ and FINEP (Brazil); NSFC (China);
CNRS/IN2P3 (France); BMBF, DFG, HGF and MPG (Germany); SFI (Ireland); INFN (Italy);
FOM and NWO (The Netherlands); MNiSW and NCN (Poland); MEN/IFA (Romania);
MinES and FANO (Russia); MinECo (Spain); SNSF and SER (Switzerland);
NASU (Ukraine); STFC (United Kingdom); NSF (USA).
The Tier1 computing centres are supported by IN2P3 (France), KIT and BMBF
(Germany), INFN (Italy), NWO and SURF (The Netherlands), PIC (Spain), GridPP
(United Kingdom).
We are indebted to the communities behind the multiple open
source software packages on which we depend. We are also thankful for the
computing resources and the access to software R\&D tools provided by Yandex LLC (Russia).
Individual groups or members have received support from
EPLANET, Marie Sk\l{}odowska-Curie Actions and ERC (European Union),
Conseil g\'{e}n\'{e}ral de Haute-Savoie, Labex ENIGMASS and OCEVU,
R\'{e}gion Auvergne (France), RFBR (Russia), XuntaGal and GENCAT (Spain), Royal Society and Royal
Commission for the Exhibition of 1851 (United Kingdom).



\addcontentsline{toc}{section}{References}
\setboolean{inbibliography}{true}
\bibliographystyle{LHCb}
\bibliography{main,LHCb-PAPER,LHCb-CONF,LHCb-DP,LHCb-TDR}

\newpage

\centerline{\large\bf LHCb collaboration}
\begin{flushleft}
\small
R.~Aaij$^{41}$, 
B.~Adeva$^{37}$, 
M.~Adinolfi$^{46}$, 
A.~Affolder$^{52}$, 
Z.~Ajaltouni$^{5}$, 
S.~Akar$^{6}$, 
J.~Albrecht$^{9}$, 
F.~Alessio$^{38}$, 
M.~Alexander$^{51}$, 
S.~Ali$^{41}$
% ...

{\footnotesize \it
% ...
$ ^{5}$Clermont Universit\'{e}, Universit\'{e} Blaise Pascal, CNRS/IN2P3, LPC, Clermont-Ferrand, France\\
$ ^{6}$CPPM, Aix-Marseille Universit\'{e}, CNRS/IN2P3, Marseille, France\\
% ...
$ ^{9}$Fakult\"{a}t Physik, Technische Universit\"{a}t Dortmund, Dortmund, Germany\\
% ...
$ ^{37}$Universidad de Santiago de Compostela, Santiago de Compostela, Spain\\
$ ^{38}$European Organization for Nuclear Research (CERN), Geneva, Switzerland\\
% ...
$ ^{41}$Nikhef National Institute for Subatomic Physics, Amsterdam, The Netherlands\\
% ...
$ ^{46}$H.H. Wills Physics Laboratory, University of Bristol, Bristol, United Kingdom\\
% ...
$ ^{51}$School of Physics and Astronomy, University of Glasgow, Glasgow, United Kingdom\\
$ ^{52}$Oliver Lodge Laboratory, University of Liverpool, Liverpool, United Kingdom\\
% ...
}
\end{flushleft}

\end{document}